\documentclass[11pt]{article}
\pdfoutput=1
\usepackage[frozencache, cachedir=minted-cache]{minted}
\usepackage{amsmath}
\usepackage{amssymb}
\usepackage{subfig}
\usepackage{xcolor}
\numberwithin{equation}{section}

\usepackage{float}
\usepackage[margin=1in]{geometry}

\newtheorem{definition}{Definition}
\newtheorem{theorem}{Theorem}
\usepackage{algorithm}
\usepackage[noend]{algpseudocode}
\usepackage{graphicx}
\NewDocumentCommand{\codeword}{v}{
\texttt{\textcolor{blue}{#1}}}
\algnewcommand{\algorithmicor}{\textbf{or }}
\algnewcommand{\algorithmicassert}{\textbf{assert }}
\algnewcommand{\Or}{\algorithmicor}
\algnewcommand{\Assert}{\algorithmicassert}

\usepackage[hidelinks]{hyperref}
\hypersetup{colorlinks=true, linktoc=all,  allcolors=magenta}

\begin{document}

\begin{titlepage}

\newcommand{\HRule}{\rule{\linewidth}{0.5mm}} 

\includegraphics[width=8cm]{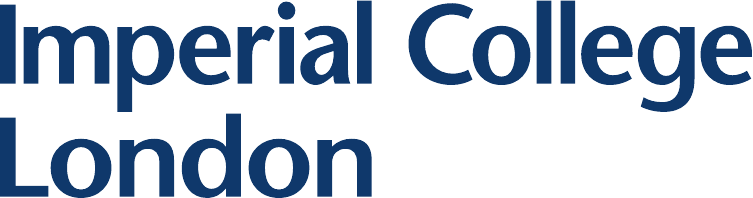}

\center 

\textsc{\LARGE Imperial College London}\\[0.5cm] 
\textsc{\Large Department of Mathematics}\\[1.5cm] 
\textsc{\Large MSci Research Project}\\[0.5cm] 

\makeatletter
\HRule \\[0.6cm]
{ \huge \bfseries Implementing a restricted function space class in Firedrake}\\[0.6cm] 
\HRule \\[1.5cm]

\begin{minipage}{0.4\textwidth}
\begin{flushleft} \large
\emph{Author:}\\
Emma Rothwell
\end{flushleft}
\end{minipage}
~
\begin{minipage}{0.5\textwidth}
\begin{flushright} \large
\emph{Supervisors:} \\
Professor David Ham, Dr Koki Sagiyama
\end{flushright}
\end{minipage}\\[2cm]
\makeatother

\vfill
Submitted in partial fulfillment of the requirements for the Mathematics Degree at Imperial College London\\[0.5cm]

\makeatletter
{\large 19/07/2024}\\[2cm] 
\makeatother

\end{titlepage}
\newpage 
\section*{Acknowledgements}\pagenumbering{gobble}
I would like to express my utmost appreciation for my supervisor, Professor David Ham. His guidance and encouragement throughout, as well as his extensive expertise in all areas of this subject, really enriched the project. \par

Additionally, I am extremely grateful for the contributions Dr Koki Sagiyama has provided to this project. His update to the code in the parallel case was invaluable for helping me overcome an extremely difficult issue. \par

I must also thank my partner Loki for all that he's done for me this year, including listening to dozens of explanations of the same concept - I hope that I have suitably returned the favour for him. In addition, my time during this project would have been much less enjoyable if I did not have the company of my friends Colman, Dak, Felix, and many others. \par

Finally, this section would be incomplete without mentioning my parents for believing in me.  
\newpage

\section*{Abstract} 
The implementation process of a \texttt{RestrictedFunctionSpace} class in Firedrake \cite{FiredrakeUserManual}, a Python library which numerically solves partial differential equations through the use of the finite element method, is documented. This includes an introduction to the current \texttt{FunctionSpace} class in Firedrake, and the key features that it has. 
\par
With the current \texttt{FunctionSpace} class, the limitations of the capabilities of the solvers in Firedrake when imposing Dirichlet boundary conditions are explored, as well as what the \texttt{RestrictedFunctionSpace} class does differently to remove these issues. These will be considered in both a mathematical way, and in the code as an abstraction of the mathematical ideas presented. \par  
Finally, the benefits to the user of the \texttt{RestrictedFunctionSpace} class are considered, and demonstrated through tests and comparisons. This leads to the conclusion that in particular, the eigensolver in Firedrake is improved through the use of the \texttt{RestrictedFunctionSpace}, through the removal of eigenvalues associated with the Dirichlet boundary conditions for a system. 

\newpage
{
\hypersetup{linkcolor=black}
\tableofcontents}
\newpage

\pagenumbering{arabic}\section{Introduction} \label{section-introduction}
Partial differential equations, or PDEs, are an important class of equations with them arising in areas such as fluids, waves, energy and diffusion. It is of great interest in numerous fields of science, from quantum physics to meteorological studies, to be able to solve these equations \cite[pp. 1]{firedrake-intro-article}. However, much like ordinary differential equations, an analytic or closed-form solution is not always known. This means that to solve a question, a researcher may be required to use numerical methods to approximate the solution of the PDE of interest. 
In general, numerical solvers will have some or all of these properties: 
\begin{enumerate}
    \item The solution must be accurate. This is the main aim of approximating the solution numerically. 
    \item The solver is efficient. In solving a PDE numerically, finer grids give a more accurate solution \cite[pp. 9]{PETSc-Numerical-Book}, but will increase the number of computations required to produce the solution. This creates the need to make sure that the method is efficient when completing large numbers of calculations. 
    \item The method has known criteria for the stability of solutions. For example, it can be shown that the simplest explicit finite difference approximation for the heat equation is stable only if $\frac{\Delta t}{\Delta x ^2} \leq 0.5$, \cite[pp. 186]{leveque} so we can choose correct spatial and time steps to ensure the error does not diverge. If we do not know much about the stability properties of the methods used, it may take a lot of trial and error, and therefore a lot of wasted time, to get a stable solution.  
    \item The method will converge to a solution. This follows from stability, it is useful for iterations to stay close to a solution if one is found. 
    \item The solver or underlying method is easy to use and understand. This could be on the software side through tutorials and proper documentation, or through the method itself. For example, a lot of finite-difference approximations come straight from Taylor series, making them quite approachable. 
    \item The method has common results through numerical analysis, such as known continuity of the solution.
    \item It is easy to store the solution in various data formats, for example for use with visualisation software such as Paraview \cite{paraview}. 
\end{enumerate}

Many methods may not fulfil all of these requirements, and some of them may not be needed for all applications. For example, a mathematician might place the most importance on requirement 6 being satisfied, while other researchers may place more emphasis on requirement 1 or 2.

The different use cases and needs, as well as numerous advances in the field, has led to a wide array of numerical methods that are currently in use for solving PDEs, each with their own advantages and disadvantages. For example, for an easy-to-use example, an explicit finite difference method is incredibly efficient at solving equations, but accuracy and stability may be sacrificed in the process. This report will focus solely on the finite element method, but it is useful for the reader to be aware that there are other approaches than are used to solve PDEs.  

The finite element method is a common method used for numerically solving partial differential equations. This involves defining a mesh, and an ``element" on each cell in the mesh. This has the advantage of being able to control continuity between cells through the use of different elements, and further numerical analysis can describe other attributes of the solution found. However, the mathematics may take longer to understand and apply, and some analysis may be too advanced for the non-mathematical user. The focus of the report is on the numerical implementation of the method through code, specifically through libraries in languages such as Python. 
\par 

A lot of finite element code is written as part of a small group of people for a specific problem \cite{firedrake-intro-article}, but this may be complex or time-consuming to code. This is where purpose-built finite element libraries are used instead, reducing the concern of incorrect code, so users can focus on obtaining a numerical solution. Currently, there are many libraries available to apply the finite element method to problems. Some are general and can solve many types of problems, while some are more niche and perhaps have features that are used more extensively in certain fields of research than others. As an example, MATLAB, a scientific computing programming language, has finite element functionality built into its Partial Differential Equations Toolbox \cite{matlab}. 
There are also numerous other libraries available, and this report will exclusively use Firedrake, as that is the one that is being modified. 
\par

Firedrake solves partial differential equations which are defined in the Unified Form Language, through the use of the finite element method. UFL is a library which is used to represent the weak formulation of partial differential expressions in code \cite{ufl}. 
The aim of Firedrake is to provide users with a ``high productivity way to create sophisticated high performance simulations" \cite[pp i.]{FiredrakeUserManual}. 
In other words, the goal of the creation and continual updating of Firedrake could be described simply as the creators want to enable the best possible user experience. With this in mind, Firedrake is always looking for ways to improve its code base, to continue to deliver a good user experience. 

For example, Firedrake uses algorithms from PETSc \cite{petsc-web-page} to ensure that large meshes are managed efficiently \cite{mesh-management-paper}, speeding up computation of various problems. Additionally, the improvements in performance and user experiences are also done by obscuring most of the computational work behind code generation and by using libraries such as PyOP2 to speed up or parallelise complex calculations. 

The result of these design choices is that the user can simply define the mesh, boundary conditions, the PDE desired expressed as a variational form, and the function space solutions should lie in and call a solver function. This takes very few lines of code, and the user only has to specify the core concepts. More advanced options exist such as using preconditioners on the system to be solved, but most functions in Firedrake only require knowledge of the finite element method and some understanding of the problem that the user wants to solve to be able to use them effectively. Additionally, there are easily accessible documentation pages and tutorials for the user to read should they require further information \cite{FiredrakeUserManual}.
\par

The contributions made to Firedrake to improve the process of solving a system with boundary conditions applied will be detailed in this report. The report  focuses on the implementation of a new class in Firedrake, designed to remove issues associated with the current methods of using function spaces on eigenvalue problems. This class also has additional benefits when solving variational problems. This will hopefully remove various non-mathematical “tricks” that Firedrake uses in the code to get around the current limitations in its implementation. As the finite element method has a lot of mathematical background, it is reasonable to assume that the authors of libraries implementing the finite element method want to reflect the mathematics behind the method as much as possible. As an extreme optimistic example, the code could be structured such that it looks entirely the same as if it were written on paper, but this is not always feasible or realistic. 
\par

The report is structured such that the definition of the finite element method is examined, and then put into Firedrake as an example of solving a variational problem. With this background in mathematics, the motivation for the changes to Firedrake is then developed, as well as some general concepts the reader should be aware of when transforming the mathematics into code. This leads into the development of the ideas of the new contributions to Firedrake. Finally, the results of these new contributions are compared to results not using them, to show the validity and utility of the class.

As Firedrake is an actively developing library, with multiple commits made to the main repository per week, some of the ideas and algorithms referenced in this paper may be subject to change or removal. For the reference of the reader, one version of Firedrake in which the changes described in the report appear in full is shown in reference \cite{my-version-of-firedrake}.
Additionally, some extra code created by the author to produce plots such as Figure \ref{bcshift-2-0-problem!} can be found at \cite{my-firedrake-supporting-code}.

The updates being regularly added to Firedrake causes additional challenges throughout the project, with the need to update the working branch every so often to the latest version, while continuing to keep changes made to the main branch. However, this need for continuous integration also means that the contributions made in this report fit in seamlessly in the wider framework of Firedrake, allowing them to be used without disruption to other areas. It also means that there is a need for clear documentation for these contributions, which this report will form a small part of. However, the main goal of the report is to convince the reader of the necessity of the contributions made to Firedrake and the reasoning behind certain choices made throughout the implementation. 

\section{Introduction to the Finite Element Method} \label{section-fem}
The finite element method (shortened in this report as FEM), as mentioned in the introduction, is one of many techniques used to solve a partial differential equation. Throughout, a general PDE in $n$ variables $x_1, ..., x_n$ is defined below \cite[pp. 2]{s-salsa-pdes-book}. 
\begin{equation}
    F\left(x_1, ..., x_n, u, \frac{\partial u}{\partial x_1}, ...,  \frac{\partial u}{\partial x_n},  \frac{\partial^2 u}{\partial x_1^2},  \frac{\partial^2 u}{\partial x_1 \partial x_2}, ...,  \frac{\partial^2 u}{\partial x_n^2},  \frac{\partial^3 u}{\partial x_1^3}, ...\right)  = 0
\end{equation}

We will also focus on Dirichlet boundary conditions that can be applied to a PDE. These can be defined as setting the value of $u$ on the boundary - in 1 dimension the boundary would be the endpoints of the interval the problem is defined on \cite[pp. 25-26]{s-salsa-pdes-book}. In a general case where $\Omega \subseteq R^{n}$, this means a Dirichlet boundary condition takes the form:
\begin{equation}
    u(\mathbf{x}) = f(\mathbf{x}) \ \ \mathbf{x} \in \partial\Omega
\end{equation}
A PDE can have multiple boundary conditions applied, affecting different areas of the domain. PDEs can also have Neumann boundary conditions imposed on them, specifying the value of a derivative of $u$ at parts of the boundary. We can also create a distinction between different values of the boundary condition:
\vspace{-0.2cm}
\begin{definition}
    Homogeneous Boundary Condition: $f(\mathbf{x}) = 0$ 
\end{definition}
\vspace{-0.7cm}
\begin{definition}
    Inhomogeneous Boundary Condition: $f(\mathbf{x}) \ne 0$ 
\end{definition}
\vspace{-0.2cm}

Now that a PDE and potential boundary conditions are defined, we turn our attention to the solution of a PDE. Some PDEs, such as the wave equation, can be solved analytically, but as mentioned in Section \ref{section-introduction}, most cannot. This is where the FEM would typically be used to attempt to solve a problem numerically rather than analytically. To understand the key steps of the FEM, first we should define a finite element. We will define a finite element using Brenner and Scott's interpretation of the original definition given by Ciarlet \cite{brenner-and-scott}:
\begin{definition}
If we have the following three items
    \begin{itemize}
        \item[i)] $K \subseteq \mathbb{R}^n$ is a bounded closed set with non-empty interior and piecewise smooth boundary
        \item[ii)] $\mathcal{P} \text{ is a finite-dimensional space of functions on } K$ 
        \item[iii)] $\mathcal{N} = {N_1, N_2, \hdots, N_k} \text{ is a basis for } \mathcal{P}'$, the dual space of P
    \end{itemize}
    Then the triplet of $(K, \mathcal{P}, \mathcal{N})$ is a finite element.
    \label{finite-element}
\end{definition}
The third item is equivalent to the condition of $N_i(v) = 0 \ \ \forall i \implies v \equiv 0$ \cite[pp. 14]{finite-element-course}. This is referred to as $\mathcal{N}$ determining $\mathcal{P}$. 
We may refer to $\mathcal{P}$ as the space of shape functions and $\mathcal{N}$ as a set of nodal variables. We can also form the nodal basis $\{\phi_i\}^{n}_{i=0}$ of $\mathcal{P}$. This is defined as the basis of $\mathcal{P}$ that is dual to $\mathcal{N}$, in other words $N_i(\phi_j) = \delta_{ij}$ \cite[pp.70]{brenner-and-scott}.

\begin{figure}[h]
    \centering
    \includegraphics[trim={8cm, 3.5cm, 7.5cm, 4cm}, clip, scale=0.3]{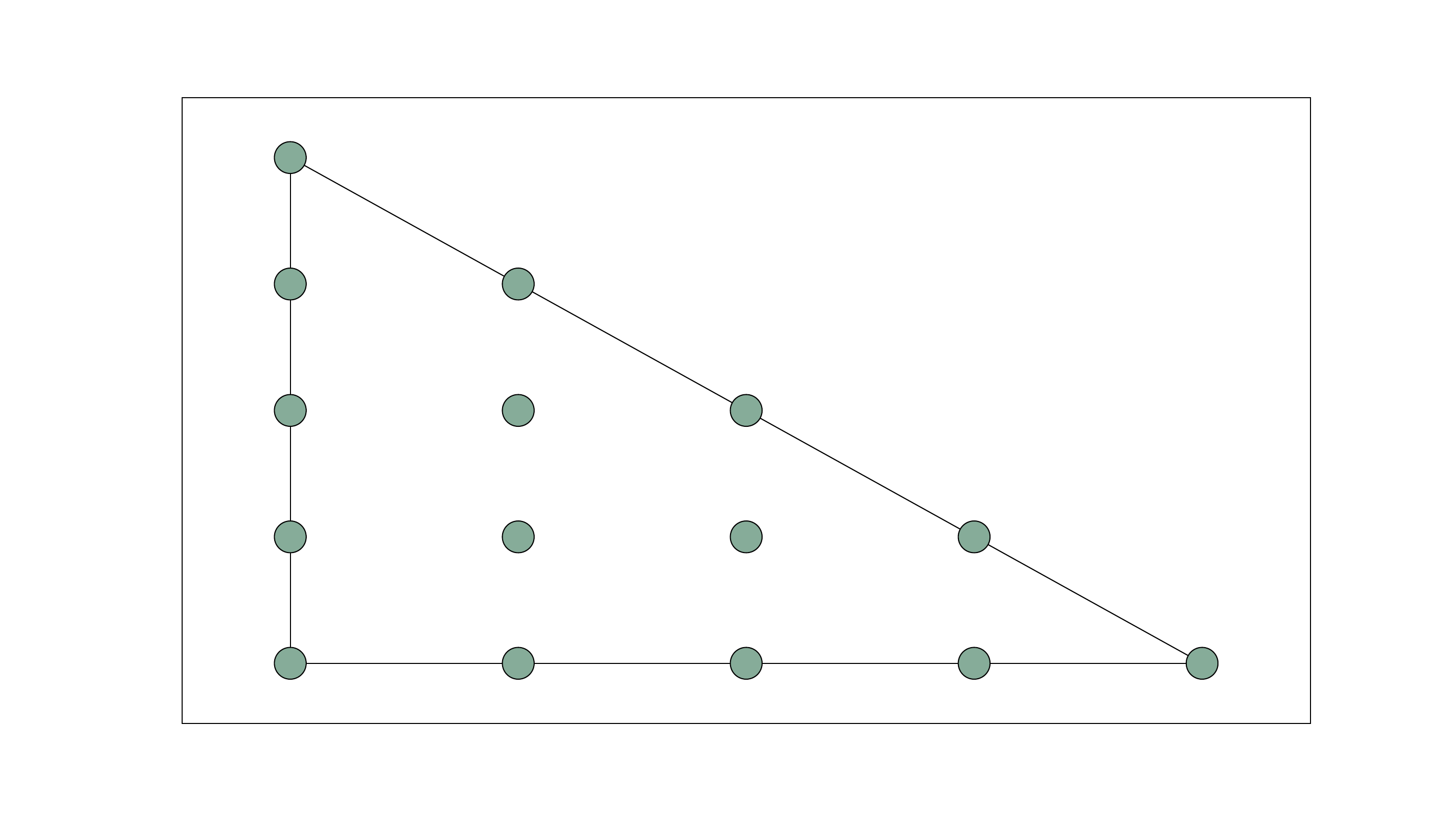}
    \caption{The Lagrange element on a reference triangle, using degree 4 polynomials. The dots represent point evaluation of $u$. \vspace{-0.3cm}}
    \label{lagrange-element}
\end{figure}

There are numerous finite elements for a variety of uses, with most of the common elements found on reference websites such as DefElement \cite{def-element}. As one of the simpler examples, we will look at the Lagrange element of degree 4. This is represented in the diagram in Figure \ref{lagrange-element}. 
\par
While in this element $K$ is a right-angled triangle, elements can be defined on other shapes as well such as intervals or quadrilaterals. We will call $K$ the reference triangle, as it displays one element and is a regular shape. This reference triangle may be stretched and transformed to different triangles if required to fit into a mesh representing $\Omega$. $\mathcal{P}$ is the space of degree 4 polynomials, and $\mathcal{N}$ is point evaluation at each of the evenly spaced points on the triangle, with $15$ points in total. The points are also known as nodes. We can check that the third condition in Definition \ref{finite-element} holds by checking that $\mathcal{N}$ determines $\mathcal{P}$. 

To find the nodal basis $\{\phi_i\}^{n}_{i=0}$ for a particular element, we can solve the system $V\mathbf{\phi} = \mathbf{\psi}$ through inversion of $V$. $V$ is the Vandermonde matrix $V_{ij} = N_{j}(\psi_{i})$, where $\psi$ is another basis of $\mathcal{P}$, such as the monomial basis. As seen in the $N_{j}(\phi_{i}) = \delta_{ij}$ equation, the nodal basis allows us to assign a unique basis function to each node in the mesh with no basis function holding weight in other nodes. 
\par

Given this example of a finite element, we can now examine the general steps we would use to solve a PDE using the finite element method on a given mesh. These are: 
\begin{enumerate}
    \vspace{-0.1cm}
    \item Determine a finite element to use, and a way of splitting the mesh that the problem is defined on into many individual elements. If the element is a triangle, this is known as a triangulation.
    \vspace{-0.6cm}
    \item Transform the partial differential equation to the weak form, through multiplication by a test function $v$ and integrating by parts. This test function should satisfy any Dirichlet boundary conditions applied to the problem. 
       \vspace{-0.1cm}
    \item Expand the trial ($u_h$) and test functions ($v$) in the nodal basis $\{\phi_i\}^{n}_{i=0}$ of the finite element chosen.  
       \vspace{-0.1cm}
    \item Assemble a matrix-vector system, formed by this expansion. This commonly is of the form $Kw = f$ where $w$ is the weighting of the basis functions. This gives the finite element approximation of the function. 
       \vspace{-0.1cm}
    \item Solve for $w$, and expand for $u_h = \sum \phi_{i}w_{i}$.
       \vspace{-0.2cm}
\end{enumerate}

Of course, there are some variations of this method depending on what problem is being solved, but this is a useful scheme to follow. This report omits some details on various aspects of the numerical analysis that comes with the method, in favour of looking at the numerical implementation. 
\par

To demonstrate the steps of the method through an example, we solve the Poisson equation on a unit square, our $\Omega$, with a homogeneous Dirichlet boundary condition on the left and right of the square. This entire example is based on the Poisson equation example from \cite[pp. 3-11]{finite-element-course}, using a different degree for the Lagrange element and some slightly different notation at times.
The full Poisson equation with boundary conditions is defined as:
\begin{equation}
    \label{poisson-equation}
    - \nabla^2 u = f, \ \ u(0, y) = u(1, y) = 0, \ \ \frac{\partial u}{\partial y}(x, 0) =  \frac{\partial u}{\partial y}(x, 1) = 0 
\end{equation}
\vspace{-0.65cm}
\begin{figure}[h!]
    \centering
    \includegraphics[trim={8.5cm, 4.3cm, 7cm, 4.5cm}, clip, scale=0.43]{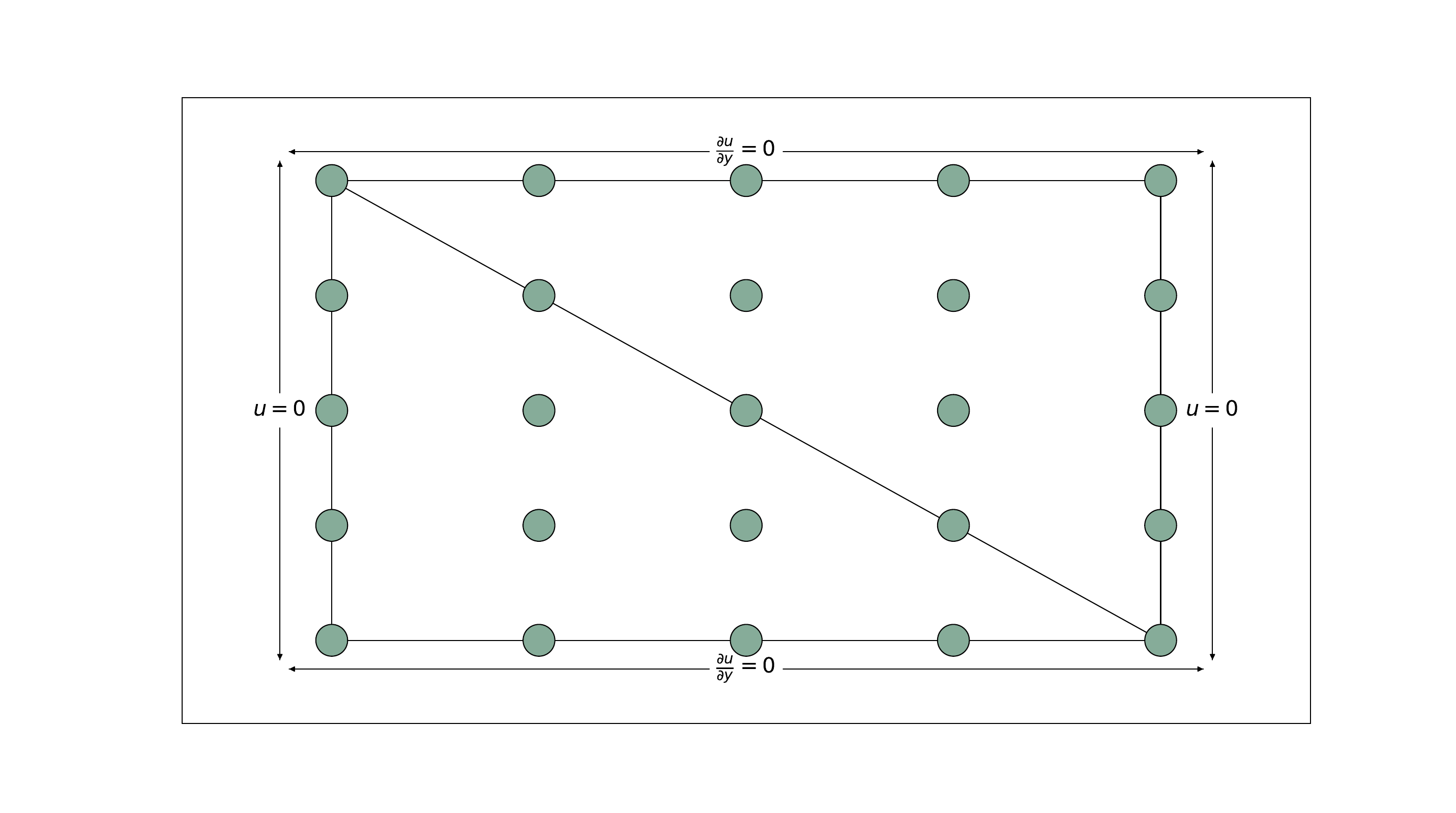}
    \caption{The unit square of two Lagrange degree 4 elements, with boundary conditions labelled. \vspace{-0.7cm}}
    \label{poisson-picture}
\end{figure}

We first triangulate our unit square in the simplest way by putting two of our Lagrange elements from Figure \ref{lagrange-element} together, to form the mesh. This triangulation, with the boundary conditions labelled, are shown in Figure \ref{poisson-picture}. The set of triangles in our triangulation will be represented as $T = \{T_1, T_2\}$.

 As in the previous figures, the green dots on Figure \ref{lagrange-element} are the nodes of the mesh. At each node, we define a set number of degrees of freedom. As we are using an element from the Lagrange family, and the nodal variables are point evaluation rather than gradient evaluation, we only have one degree of freedom for each node. 
\par

The choice in finite element dictates the choice in function space in which to solve our problem. As we are using a degree 4 element, this space will be related to the space of degree 4 polynomials. We use the function space $V$ which is defined below:
\begin{equation*}
    V = \{g : g|_{Ti} = \text{a quartic polynomial} \land g\in C^{0}(\Omega)\}
\end{equation*}
In other words, the condition that the solution is a polynomial only applies on each triangle, and there is continuity of the solution over the mesh. 
To be able to solve for $u_h$ on each triangle and to put everything together to create a solution, we need to transform the PDE to the weak formulation. To do this, as previously mentioned, we multiply the strong form of Poisson's equation given in Equation \ref{poisson-equation} by a test function $v$ and integrate by parts over the domain. The test function is in the space $V_0$, which contains functions vanishing on the boundary. We only need to look at this space, as the value of $u_h$ is already known on the boundary. This space is defined below. 
\begin{equation*}
    V_0 = \{f \in V | f(0, y) = f(1, y) = 0 \}
\end{equation*}
In this homogeneous case, we can also say that the Dirichlet boundary conditions are respected in $V_0$.
We also replace our function $u$ with an approximation $u_h \in V_0$, as $u_h$ can be expanded in the nodal basis for $V_0$ whereas $u$ was a generic function that may not even be in $V$ or $V_0$. 
This is done below, with care being taken to expand the integral over each triangle in the mesh.
\begin{align}
    \int_{\Omega} -\nabla^2u_h v d\mathbf{x} &= \sum_{i=1}^{2}\left(\int_{T_i} \nabla v \cdot \nabla u_h d\mathbf{x} - \int_{\partial T_i} v n\cdot \nabla u_h dS\right)  \nonumber \\
    &= \int_{\Omega} \nabla v \cdot \nabla u_h d\mathbf{x} \label{poisson-weak-form}  \\ 
    \implies \int_{\Omega} \nabla v \cdot \nabla u_h d\mathbf{x} &= \int_{\Omega} vf d\mathbf{x} \label{poisson-weak-form-2}
\end{align}
 As we have the same integrand being integrated on a path going both up and down the central diagonal of the square, these interior integrals cancel out. We also obtain the second equality from utilising the Neumann boundary condition over the boundary integrals. Finally, the third equality comes from the strong formulation of the PDE. These integrals can be rewritten to match the following definition \cite[pp. 33]{finite-element-course}:
 \begin{definition}
 \label{linear-var-problem-def}
     Linear Variational Problem: Find u such that a(u, v) = G(v) $\forall v \in V$, where $a(u, v)$ is a bilinear form and $G(v)$ is a linear form.
 \end{definition}
Further definitions on variational problems exist, expressing more details such as whether or not a variational problem is symmetric, as seen in \cite[pp. 57-58]{brenner-and-scott}. However, in the report only a distinction between linear and non-linear problems is necessary, so these definitions will not be explored.
In general, we write the variational problem on $u$, and then derive the finite element approximation, $u_h$ for $u$. In the case of the Poisson problem, this means that we want to solve $a(u, v) = \int_{\Omega} \nabla v \cdot \nabla u \ dx, \ G(v) = \int_{\Omega} vf dx$. 

A lot of numerical analysis of the finite element method can come from the study of properties of the components of the linear variational problem. For example, the following is required to show that a unique solution does exist through the Lax-Milgram theorem \cite[pp. 62]{brenner-and-scott}, which may be one of the first properties considered when formulating and solving a problem. 

\begin{theorem}
Lax-Milgram Theorem: If $a(u, v)$ is both continuous and coercive, and $G(v)$ is coercive, then there exists a unique $u$ solving the variational problem $a(u, v) = G(v) \forall v \in V$.     
\end{theorem}

For example, the Poisson equation as expressed in the variational form given in Equation \ref{poisson-weak-form-2} can be shown to have a unique solution. Furthermore, it can be shown that under certain conditions on $f$, a solution to the variational problem in \ref{poisson-weak-form-2} also is the solution for the original partial differential equation \cite[pp. 131]{brenner-and-scott}. Therefore, it is reasonable to expect to be able to solve the problem using the finite element method. 

From the variational form from Equation \ref{poisson-weak-form-2}, we have two integrals which must be discretised using the nodes on the mesh, so that we can form a system to be solved. This is done through expansion of $u_h$ and $v$ in the nodal basis $\phi$ of $V_0$, corresponding to Step 3 in the list. As there are two Dirichlet boundary conditions, one on each of the left and right boundaries, we remove the basis functions corresponding to nodes from the left and right-hand side of our square as our function in $V_0$ vanishes on the boundary. This gives $15$ basis functions for $V_0$, compared to $25$ basis functions for $V$. This change is reflected in the following equations, as the sum is from $0$ to $14$ instead of $0$ to $24$. 
\begin{equation}
u_h = \sum^{14}_{i=0} u_i \phi_i,  \  \ v = \sum^{14}_{i=0} v_i \phi_i, \ \  v_i, u_i \in \mathbb{R} 
\end{equation}
Substituting these expansions into Equation \ref{poisson-weak-form-2}, we get:
\begin{equation}
       \int_{\Omega} \sum^{14}_{i=0} v_i \nabla \phi_i \cdot  \sum^{14}_{i=0} u_i \nabla \phi_i  \ d\mathbf{x} =  \int_{\Omega} \sum^{14}_{i=0} v_i \phi_i f d\mathbf{x}
\end{equation}
Rearranging to remove the sum in $v_i$, we obtain the following:
\begin{equation}
       \sum^{14}_{i=0}  v_i  \left( \int_{\Omega}  \sum^{14}_{j=0} u_j \left( \nabla \phi_i \cdot \nabla \phi_j \right) \ d\mathbf{x} -  \int_{\Omega}  \phi_i f d\mathbf{x} \right) = 0 
\end{equation}
This shows that the system does not depend on the coefficients $v_i$, reflecting the fact the weak formulation must hold for any test function. 
We can now assemble the matrix-vector system $Ku = f$, defined as:
\begin{equation}
    f_i = -  \int_{\Omega}  \phi_i f \ d\mathbf{x}, \ \ u_i = u_i \ \ K_{ij} = \int_{\Omega} \nabla \phi_i \cdot \nabla \phi_j d\mathbf{x} 
    \label{matrix-vector-system}
\end{equation}
 To solve for our approximation for $u_h$, we would then need to invert the matrix K, or use a similar operation if inversion is impossible. As we can see, this matrix is of size $15\times15$, far too big for manual inversion. Additionally, if our mesh was more complex than just two elements put together, or we used a different element with more basis functions on the mesh, we might not know how to uniquely identify each node to keep consistency in our solution. Due to these computational requirements, there are now a wide variety of finite element solvers that help the end user with almost every stage of the process. Firedrake is one such package, the usage of Firedrake to solve such problems is the focus of the next section.

\section{Introduction to Firedrake}\label{section-firedrake-intro}
As mentioned in Section \ref{section-introduction}, Firedrake is a Python library that numerically solves partial differential equations, defined by the user through use of the Unified Form Language (UFL). Firedrake supports many of the finite elements that are commonly used to solve problems, such as the Lagrange element defined in Section \ref{section-fem} or more uncommon ones such as the Hermite elements. 

The focus of this section is on the application of Firedrake to solving a PDE, although other functionality such as solvers for eigenproblems come with the package. These solvers will be discussed in more detail in later sections, as one of the main motivations for the creation of the \texttt{RestrictedFunctionSpace}. 

The lines of code used to solve a problem in Firedrake are intentionally similar to the list of steps required in \ref{section-fem}. The only objects that need to be defined are the mesh, the \texttt{FunctionSpace} as a representative for the finite element, any boundary conditions on the problem and the forms for the variational problem as expressed in UFL.

The use of UFL to express the variational problem allows users to write $\int \nabla u \cdot \nabla v \ dx$ in Python as \texttt{inner(grad(u), grad(v)) * dx}. This is immediately recognisable to the person who wrote the problem, so when using UFL it is easy to translate the mathematics to Python code. The obvious benefit of this is that the user does not have to learn lots of syntax to define their problem using UFL, allowing for less time required to create code for a solution to a problem. 

To look at how Firedrake is used to solve the Poisson equation example given in Section \ref{section-fem}, replicating the steps in the list, we created a minimal code example when $f = x^2$, reproducible via Listing \ref{poisson-code}.

\begin{listing}[h!]
\begin{minted}{Python}

from firedrake import *
mesh = UnitSquareMesh(1, 1)
x, y = SpatialCoordinate(mesh)
# create Lagrange element of degree 4
V = FunctionSpace(mesh, "Lagrange", 4)

u = TrialFunction(V)
v = TestFunction(V)

# 1 indicates the left side of mesh, 2 for right side
bc_left = DirichletBC(V, 0.0, 1)
bc_right = DirichletBC(V, 0.0, 2)

# create bilinear form, dx means integral
a = inner(grad(u), grad(v)) * dx

# create linear form
f = Function(V).interpolate(x**2)
G = inner(f, v) * dx 

# solve, u_sol holds solution
u_sol = Function(V)
solve(a == G, u_sol, bcs=[bc_left, bc_right])
\end{minted}

\caption{Code for solving the Poisson problem of $-\Delta u = x^2$ with 2 homogeneous Dirichlet boundary conditions on the unit square. This code uses degree 4 Lagrange elements, as seen in Section \ref{section-fem}.}
\label{poisson-code}
\end{listing}

The solution that results from Listing \ref{poisson-code} has $25$ data points, one for each node. This is verified by looking at the value of \texttt{len(u\_sol.dat.data)}. No assembly of the matrix-vector system such as Equation \ref{matrix-vector-system} is ever seen by the user when \texttt{solve} is called, although this does happen inside the code. If the user is interested in returning the assembled matrix on the left-hand side, Firedrake has a function called \texttt{assemble} that can be used instead. 

The report will look at how the treatment of boundary conditions, specifically Dirichlet boundary conditions, is implemented in Firedrake in Section \ref{section-firedrake-bcs}. Later sections in the report will focus on Firedrake again, specifically on the implementation of the \texttt{FunctionSpace} class, and what it represents.

\newpage
\subsection{Treatment of Boundary Conditions}\label{section-firedrake-bcs}
The manual for Firedrake has a section dedicated to the discussion of how Dirichlet boundary conditions are treated both mathematically and in the code \cite[pp. 27-31]{FiredrakeUserManual}, and how the function space can be split in accordance with the boundary conditions. A detailed look at this manual section will be presented in this section, as the treatment of boundary conditions is an important feature that changes between the ideas implemented in this project and the current Firedrake code. This entire section naturally closely follows the ideas developed and presented in the manual \cite[pp. 27-31]{FiredrakeUserManual}. This section only presents the ideas that are in Firedrake currently, and in Section \ref{section-motivation} the issues arising from this approach are discussed. 

In the manual chapter, it is seen that the method of solving variational problems in Firedrake is the same for both linear and non-linear variational problems. We have not studied a non-linear variational problem before in this report, but the manual defines such a problem to have the form of $F(u; v) = 0$. This form of a non-linear variational problem is referred to as the residual form, which is linear in $v$ but not necessarily in $u$.
From Definition \ref{linear-var-problem-def}, it is clear that a linear variational problem can always be rewritten in this form, by simply setting $F(u; v) = a(u, v) - G(v)$.

When imposing a Dirichlet boundary condition of $u(x) = g(x) \ \ x \in \Gamma$, which can be either homogeneous or inhomogeneous, we can split the function space that the problem is defined on into two parts. 
\begin{equation}
    \label{fs-decomposition}
    V = V_{0} \oplus V_{\Gamma}
\end{equation}

$V_{0}$ is spanned by the basis functions $\phi_i$ that vanish on the boundary $\Gamma$ and $V_{\Gamma}$ is spanned by all the other basis functions. $V_0$ is the same space as $V_0$ from Section \ref{section-fem}. Functions in $V_{\Gamma}$ are the functions used to ensure the boundary conditions hold in the solution. From these spaces, we label the set of basis functions of $V_\Gamma$ as $\mathbf{\phi}^\Gamma$, and use the label $\mathbf{\phi}^0$ for the basis functions of $V_0$.

Taking this decomposition, as we know that $v \in V$, we can use the linearity of $F$ in $v$ and split up the non-linear variational problem into two components. One component takes the $v$ argument as $v_0 \in V_0$ and the other as $v_\Gamma \in V_\Gamma$, using the decomposition of $v = v_0 + v_\Gamma$. More clearly, this is written as the following:
\begin{equation}
    F(u; v) = F(u; v_0) + F(u; v_\Gamma)
\end{equation}
Using this split form of the residual, we will define that the term in $v_\Gamma$ is 0 when $u$ satisfies the boundary condition. This is because our residual $F(u; v_\Gamma)$ measures a deviation from the solution to the variational problem, but there are fixed values on the boundary which could clash with this depending on the values of the boundary condition. This is resolved when the residual is $0$, when the correct value is found on the boundary. 

The second function in the definition of the residual is now denoted as $F_\Gamma$, defined below, where $g_i$ is the evaluation of $g(x)$ at the $i^{th}$ node. 
\begin{equation}
    F_\Gamma(u; \phi_i) = u_i - g_i, \ \phi_i \in \phi^\Gamma
\end{equation}
This then gives us the following equation for our residual.
\begin{equation}
    \hat{F}(u; v_0 + v_\Gamma) = F(u; v_0) + F_\Gamma(u; v_\Gamma) = 0 \ \ \forall v_0 \in V_0, v_\Gamma \in V_\Gamma
\end{equation}

This equation is solved by Newton's method, using one iteration if the problem is linear and potentially multiple iterations required if nonlinear. Below is the standard Newton update step:
\begin{equation}
    \label{newton-iteration}
    U_{new} = U_{old} - J^{-1}F(u) \ \ \ F(u)_{i} = \hat{F}(u; \phi_i), \ \ J_{ij} = dF(u; \phi_{i}, \phi_{j})
\end{equation}
If we use this definition of $J_{ij}$ alongside the definition of the derivative, we can see the behaviour on boundary rows when $\phi_{i}\in V^{\Gamma}$:
\begin{align*}
    dF(u; \phi_{i}, \phi_{j}) = \text{lim}_{h \rightarrow 0}\frac{\hat{F}(u + h\phi_i; \phi_j) - \hat{F}(u; \phi_{j}}{h} \\ 
    \phi_{j} \in V^\Gamma \implies  dF(u; \phi_{i}, \phi_{j}) = \text{lim}_{h \rightarrow 0} \frac{F_\Gamma(u + h\phi_i; \phi_j) - F_\Gamma(u; \phi_j)}{h} \\
    F_\Gamma(u + h\phi_i; \phi_j) = (h\phi_i + \sum u_k \phi_k)_j  - g_j = (u_j)(1 + h\delta_{ij}) - g_j \\ \implies \frac{F_\Gamma(u + h\phi_i; \phi_j) - F_\Gamma(u; \phi_j)}{h} = \delta_{ij} \\ 
    \phi_{j} \notin V^{\Gamma} \implies dF(u; \phi_{i}, \phi_{j}) = \text{lim}_{h \rightarrow 0} \frac{F(u + h\phi_i; \phi_j) - F(u; \phi_j)}{h} = 0 \\ 
   \implies J_{ij} =  \delta_{ij}, \ \ i = j, \phi_i \in V^{\Gamma} 
\end{align*}

The value for this derivative means that the rows of the Jacobian corresponding to nodes affected by boundary conditions are rows of the identity matrix, making them easy to identify when observing the Jacobian matrices created during the update step.

Using this definition of the Jacobian, we can see that it can be decomposed into 4 blocks. This is valid for all arrangements of the basis functions of $V$, as they can be relabelled to obtain the arrangement of ${\mathbf{\phi}^\Gamma, \mathbf{\phi}^0}$. 
\begin{equation*}
    J = \begin{pmatrix}
        J^{\Gamma \Gamma} & J^{\Gamma 0 } \\ 
        J^{0 \Gamma} & J^{00}
    \end{pmatrix}
\end{equation*}

The superscripts are used to refer to the space of the basis function associated with the row/column. For example:
\begin{equation*}
    J^{0\Gamma}_{ij} = \begin{cases}
        J_{ij} \ \  \phi_i \in \phi^{0}, \phi_{j} \in \mathbf{\phi}^{\Gamma} \\
        0 \ \ \text{otherwise}
    \end{cases}
\end{equation*}
$J^{\Gamma 0} = \mathbf{0}$ as in this sub-block we can see the rows of the sub-block are the boundary rows when $\phi_i \in V^\Gamma$, but we know $i \ne j$ as $\phi_j \in V^{0}$. Using similar reasoning, $J^{\Gamma\Gamma}$ is the identity matrix of size $|\mathbf{\phi}^\Gamma|$.

If we represent the update of one Newton step with $\hat{U} = U_{old} - U_{new}$, we can see from Equation \ref{newton-iteration} that we will need to solve the following to obtain the value of $\hat{U}$:
\begin{equation*}
    J\hat{U} = F(u)
\end{equation*}

In the boundary rows, it has been shown that the Jacobian is an identity row, so we can create a separate update called $U^\Gamma$ for the case where $\phi_i \in \phi^\Gamma$:
\begin{equation*}
    U^{\Gamma}_{i} = \begin{cases}F(u)_{i} \ \ \phi_i \in \phi^{\Gamma} \\ 
    0 \ \ \text{otherwise}
    \end{cases}
\end{equation*}

Putting everything together, using forward substitution, we can see that the system $J\hat{U}= F(u)$, is equivalent to:
\begin{equation}
\label{final-update-newton-J-decomposition}
    (J^{00} + J^{\Gamma\Gamma})\hat{U} = F(u) - J^{0\Gamma}\hat{U}^{\Gamma}
\end{equation} 

The advantage of this split of the residual is that if the initial guess for the solution provided to the solver is correct on the boundary, in the sense that the Dirichlet boundary conditions imposed on the system are respected, we have that the update value on the boundary rows will be $0$. This means these correct boundary guesses will not be updated, while the solution in the rest of the nodes are updated and will converge to the correct solution. Of course, some users may input an initial guess that does not take the correct value on the boundary. A common example of this is initialising a function that takes the value of $0$ everywhere, while having inhomogeneous Dirichlet boundary conditions. In the \texttt{solve} function in Firedrake, to avoid the possibility of a non-zero update on the nodes with boundary conditions, all supplied Dirichlet boundary conditions are applied to the initial guess to manually set the values to be as expected before solving using Newton iteration takes place.

\section{Motivation for a Restricted Function Space}\label{section-motivation}
Given this description of the treatment of Dirichlet boundary conditions in Firedrake, it is a good idea to question what possible issues occur from this and why they might occur. This will provide motivation for the \texttt{RestrictedFunctionSpace} class that is produced in this report.

When formulating and solving a variational problem in Firedrake, Dirichlet boundary conditions and where on the mesh they are applied to are only introduced to the problem when \texttt{solve} is called, rather than at the initial definition of the function space that the solution will be contained in. Mathematically, and indeed in Firedrake, this has the effect of also taking trial and test functions from the overall function space $V$, rather than the reduced space $V_0$, as the order of information presented means that the space of solutions is set before boundary conditions are defined on the specific variational problem. This means that the variational problem is solved for functions in $V$ rather than for functions in $V_0$ with a reintroduction of the solution of the boundary.

This can be seen through the derivations of Section \ref{section-firedrake-bcs}, as the Jacobian block $J^{\Gamma\Gamma}$ is present in the Newton update step given by Equation \ref{final-update-newton-J-decomposition}. In our Poisson example, this is a difference between a $25 \times 25$ matrix and a $15 \times 15$ matrix assembled and solved in our matrix-vector system, with just two cells. If the mesh is substantially more complex, or if a higher-order element is used, the size difference between the matrices may become larger. This efficiency problem has also been considered in other works, such as \cite[pp. 71-72]{matt-knepley-book}. In this, the author states that due to the fact that code used to implement these Dirichlet boundary conditions, in the same manner as Firedrake, has additional steps such as the setting of boundary values there are challenges in optimising the code.

Additionally, when a Jacobian is assembled the rows corresponding to boundary nodes are identity rows. The boundary values do not get updated throughout the solving procedure, and so it means there is a calculation of an update of $0$ at each Newton iteration for these boundary values. To ensure that the update is $0$ at each step, the entries of the nodes on the boundary conditions are set to be that of their boundary condition value by the \texttt{solve} function. Some users may be optimising an initial guess which involves the boundary nodes, where the values of the guess on the boundary are not the same as the values of the boundary condition. The movement of their guess within the function, while necessary, is not particularly transparent and may confuse users unfamiliar with the reasons as to why this occurs. A better approach would be to set the values on the boundary nodes at the start, and then not include them in the solving at all, to make the distinction clearer to the regular user between fixed values and values that require solving a system to find. 

Another, more serious, issue is the treatment of eigenvalue problems in Firedrake. Firedrake contains an eigensolver class called \texttt{LinearEigensolver} used to solve problems of the form $A\psi = \lambda M \psi$. However, in the class constructor, forms are passed in rather than the preassembled matrices $A$ and $M$, as well as a list of boundary conditions. This means $A$ and $M$ will be assembled in the code, with boundary conditions passed in. When assembling a matrix and boundary conditions are supplied, boundary degrees of freedom result in identity rows in the assembled matrix, the same as in the Jacobian. 

As the space that the problem is defined in is $V$, when boundary conditions are applied to the eigenproblem, basis functions in $\phi^\Gamma$ are still a part of the system. The consequence of this is that there are some eigenvalues associated with the space of functions satisfying the boundary condition that are returned from the solver. These are not the eigenvalues of interest, and furthermore mean that the output from the eigensolver may not allow the user to distinguish between these boundary eigenvalues and the other eigenvalues. In the worst case, the duplication of eigenvalues may lead to convergence failures. 

As a small example of an eigenproblem, here is a $3\times3$ system of the form $A\psi = \lambda M \psi$:
\begin{equation}
\label{small-eigenvalue-problem}
    \begin{pmatrix}
    1 & 0 & 0 \\ 
    0 & 0.5 & 1 \\ 
    0 & 3 & 2 \\ 
    \end{pmatrix}\psi = \lambda \begin{pmatrix}
        1 & 0 & 0 \\ 
        0 & 0.5 & 0.25\\ 
        0 & 3 & -0.5\\ 
    \end{pmatrix}\psi 
\end{equation}

Through multiplication by $M^{-1}$, we obtain that two of the eigenvalues of this system are 1, with the third one being 2. The top row of both $A$ and $M$ is an identity row, which means this row represents a boundary eigenvalue. Without prior information on the quantity of boundary degrees of freedom, we might decide to treat both or neither of these eigenvalues of $1$ as non-boundary eigenvalues, which has effects on the corresponding eigenvectors. 

The solution that is currently used in the code of Firedrake is to shift the eigenvalues of boundary rows away from the same value as other interior eigenvalues. This is achieved by introducing a shift parameter, which we will denote by $\theta$. The $\theta$ has the following effect on $A$ and $M$ of the original system, using $i$ for rows:
\begin{align*}
    [A(\theta)]_{i} &= A_{i} \\ 
    [M(\theta)]_{i} &= \begin{cases}
        \frac{1}{\theta} M_{i} \ \ \phi_{i} \in \phi^\Gamma \\ 
        M_{i} \ \ \text{otherwise}
    \end{cases}
\end{align*}
That is to say, only the identity rows of $M$ are multiplied by $(\theta)^{-1
}$, while the rest of the eigenproblem system remains unchanged. This has the effect of shifting the value of the boundary condition eigenvalues from $1$ to $\theta$ , as on the eigenvalue rows we now solve the system $v = \frac{\lambda}{\theta} v$ for $\lambda$.

The problem with this approach is that if nothing about the system that is being solved is known, there may still be clashes. These clashes could have the consequence of meaning the solver still fails to converge if the shifted eigenvalues end up being close to the original eigenvalues.
If we use this feature, and have $\theta = 5$ in our system setting up the problem in Equation \ref{small-eigenvalue-problem}, we end up with this assembled form:
\begin{equation}
    \label{shifted-eigenvalue-problem}
    \begin{pmatrix}
    1 & 0 & 0 \\ 
    0 & 0.5 & 1 \\ 
    0 & 3 & 2 \\ 
    \end{pmatrix}\psi = \lambda \begin{pmatrix}
        0.2 & 0 & 0 \\ 
        0 & 0.5 & 0.25\\ 
        0 & 3 & -0.5 \\ 
    \end{pmatrix}\psi 
\end{equation}
Notice how the left-hand side remains the same, but the right-hand side has the first row multiplied by $1/5$. This gives the result that the first eigenvalue is now $5$, rather than $1$, and as we know the parameter is $5$ we can be reasonably certain this eigenvalue is the only one corresponding to the boundary and so is to be ignored. However, if we were to use $\theta = 2$ in the system we end up with the same issue as with the system in Equation \ref{small-eigenvalue-problem}, but instead we now have difficulty distinguishing between the two eigenvalues of value 2. This is why information about the system is required to use the shift of $\theta$, even if convergence of the eigensolver is possible.

\begin{figure}[h!]
    \centering
    \includegraphics[trim={2cm, 1cm, 2cm, 1cm}, scale=0.35]{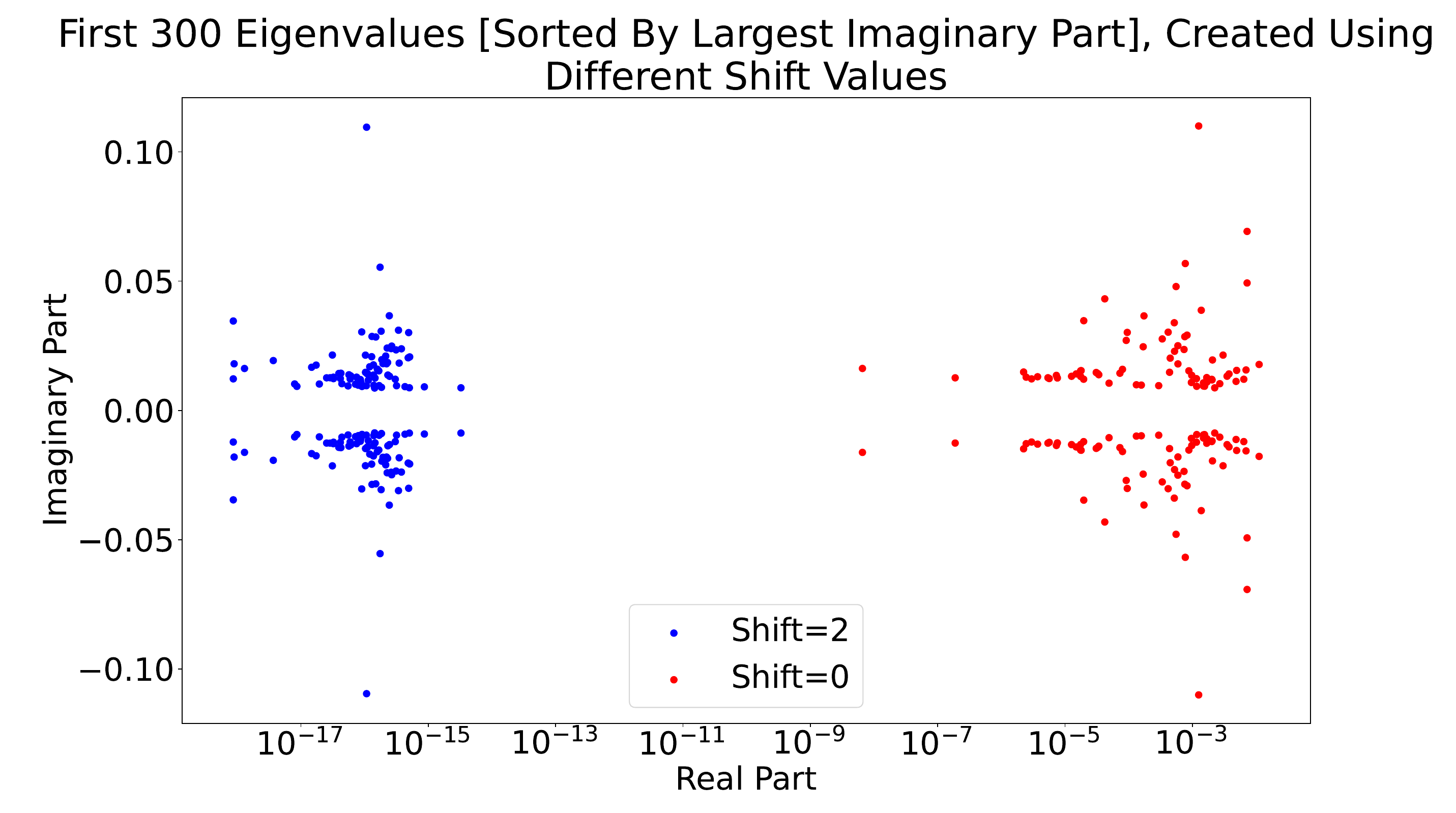}
    \caption{Scatter plot of the first 300 eigenvalues of the Oceanic basin modes problem. These eigenvalues are generated for the shift parameter $\theta = 0$ (red) and $\theta = 2$ (blue)}
    \label{bcshift-2-0-problem!}
\end{figure}

To demonstrate a problem in Firedrake where the eigensolver does not converge in the way we would have expected, we use the following code. This code is based on the tutorial for the eigensolver in Firedrake \cite[pp. 122-127]{FiredrakeUserManual}, which solves the following eigenproblem, expressed in the integral form in Equation \ref{oceanic-basins-eigenproblem}. This has a Dirichlet boundary condition of $\phi = 0$ on the entire boundary of the mesh.  
\begin{equation}
\label{oceanic-basins-eigenproblem}
    \text{Find } \psi, \ \lambda \text{ such that: }- \lambda \int \int_{A} (\nabla \phi \cdot \nabla \psi + \phi\psi)  dA  =  \int \int_{A} \phi \cdot \frac{\partial \psi}{\partial x} dA \ \  \forall \phi \in V
\end{equation}
This means the problem is of the form $A\psi = \lambda M \psi$. Listing \ref{eigenproblem-code} shows how a user could solve this eigenproblem in Firedrake, shifting the eigenvalues on the boundary by 2. \\
\vspace{-0.5cm}
\begin{listing}[h!]
\begin{minted}{Python}
from firedrake import *
Lx, Ly = 1, 1
n0 = 50
mesh = RectangleMesh(n0, n0, Lx, Ly, reorder=None)
V = FunctionSpace(mesh, "CG", 1)
bc = DirichletBC(V, 0, "on_boundary")
n = 300
phi, psi = TestFunction(V), TrialFunction(V)

opts = {"eps_gen_non_hermitian": None, "eps_largest_imaginary": None,
        "st_type": "shift", "eps_target": None,
        "st_pc_factor_shift_type": "NONZERO"} # options for solver

eigenproblem_2_shift = LinearEigenproblem(
        A=phi*psi.dx(0)*dx,
        M=-inner(grad(psi), grad(phi))*dx - psi*phi*dx,
        bcs=bc, bc_shift=2, restrict=False)
eigensolver_2_shift = LinearEigensolver(eigenproblem_2_shift, n_evals=n,
                                        solver_parameters=opts) 
nconv_2_shift = eigensolver_2_shift.solve()
\end{minted}
    \vspace{-0.5cm}
    \caption{Code to solve the eigenproblem first described in Equation \ref{oceanic-basins-eigenproblem} through Firedrake. This employs a shift parameter of $\theta$ = 2, represented as \texttt{bc\_shift} when constructing the \texttt{LinearEigenproblem} class.}
    \label{eigenproblem-code}
\end{listing} \\ 
Using this code, a plot of the eigenvalues that are returned when using two different values of the $\theta$ parameter is produced. This is shown in Figure \ref{bcshift-2-0-problem!}. 
When solving an eigenproblem with Firedrake, there are options that can be set to specify which eigenvalues should be searched for first. In the plot in Figure \ref{bcshift-2-0-problem!}, the eigensolver is finding the eigenvalues with the largest imaginary value by using the \texttt{eps\_largest\_imaginary} option within the \texttt{opts} dictionary given in Listing \ref{eigenproblem-code}. By doing this, the comparability of two eigensolver results should be improved, as without a set option the eigenvalues could be returned in a random order.

Clearly, the two clusters of eigenvalues are different, with one cluster being centred at a value numerically close to 0 and the other cluster situated closer to $10^{-3}$. Due to the proximity of the $\theta=2$ cluster to $0$, the $\theta=0$ parameter is likely an unwise choice, and is the reason for that cluster of eigenvalues moving away from zero. Additionally, the eigenvalues found are of the form $\lambda = i \omega$ \cite[pp. 122-127]{FiredrakeUserManual}, so should only contain an imaginary part. From this, it is clear that the eigenvalues for the shift of $\theta = 0$ are not accurate, as their real part is relatively large compared to the imaginary part. 

For the solution of the eigenproblem, $336$ eigenvalues are found for the problem when we use the shift of $\theta=2$, whereas only $300$ are found when the shift was $\theta=0$. This is a noticeable difference in the number of eigenvalues found, suggesting that the problem with a shift of $2$ is somehow better conditioned than the problem with a shift of $0$. This could lead to the problem not being solved at all if no eigenvalues were found, meaning if nothing about the structure of the eigenvalues is known at the time of solving there could be a lot of experimentation to find the correct shift, wasting time. 

\section{General Implementation Concepts} \label{section-general-implementation}
In this section, some of the concepts used in Firedrake and associated packages will be examined in order to describe how the implementation is carried out in later sections.

The majority of this section is about the various labels applied to a mesh from various libraries, their use, and how they can be used to rebuild the original mesh if so desired. These labels are what is changed in the new class that is introduced in the project, so it is important to ensure the reader is familiar with this terminology.

\subsection{Local and Global Numbering}\label{section-local-global}
\begin{figure}[h!]
    \centering
    \includegraphics[trim={7.5cm, 4.1cm, 6cm, 4.5cm},clip, scale=0.36]{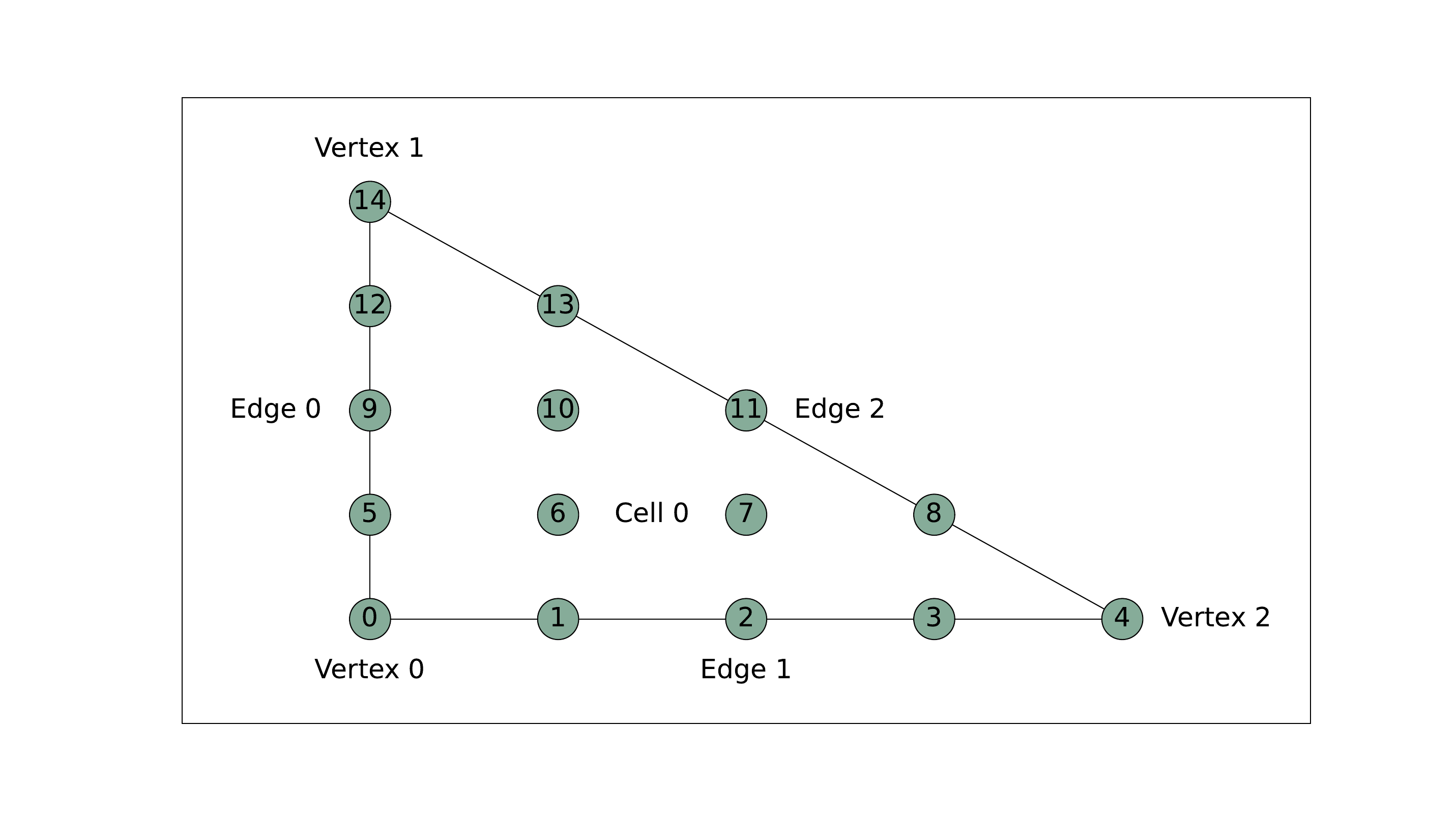}
    \caption{One possible numbering of the Lagrange element, with topological entities also numbered.\vspace{-0.5cm}}
    \label{lagrange-element-numbered}
\end{figure}

As mentioned in Section \ref{section-fem}, there is a need to organise the nodes of each element and of the entire mesh. Without a system in place, it is possible that nodes will get swapped around, and a solution is returned that does not make sense. To remove this, local and global numbering is used. We define them below \cite[pp. 87]{finite-element-course}.
\vspace{-0.3cm}
\begin{definition}
    Local Numbering: A map which relates the nodes of the finite element on a singular cell to the topological entities of the cell.
\end{definition}
\vspace{-0.5cm}
\begin{definition}
    Global Numbering: An map from the mesh entities to the number of degrees of freedom.
\end{definition}
\vspace{-0.3cm}
In the 2D case, these mesh entities are defined to be one of a vertex, an edge or an entire cell, and the definition extends appropriately to higher or lower dimensions \cite[pp. 72]{finite-element-course}.
As an example, a possible numbering of the element described in Figure \ref{lagrange-element} is shown in Figure \ref{lagrange-element-numbered}, although there are multiple possible numbering systems that could be used instead. \par

\begin{figure}[h!]
    \centering
    \includegraphics[trim={9cm, 4.6cm, 8cm, 5cm}, clip, scale=0.4]{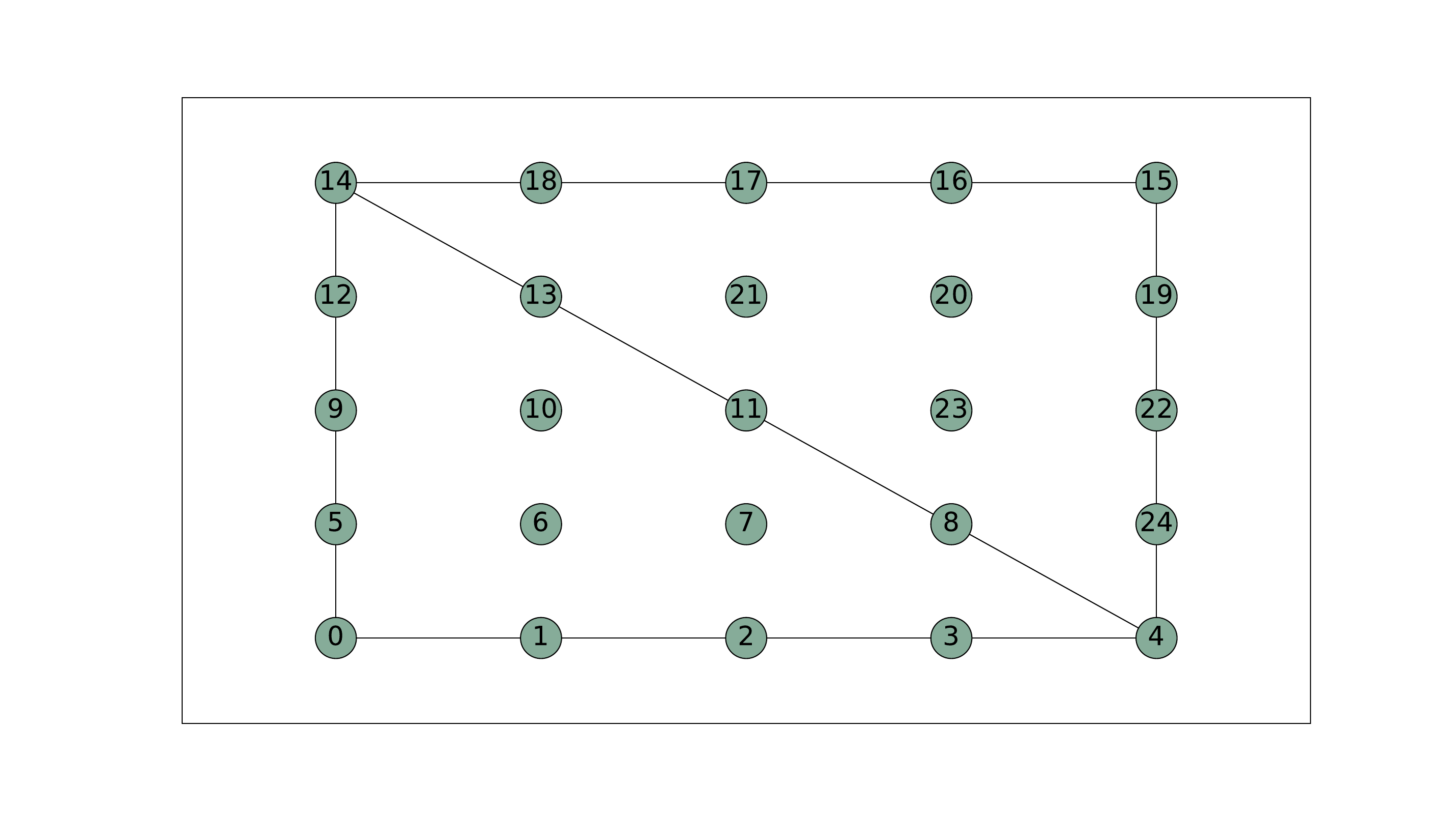}
    \caption{One possible global numbering of the problem described in Figure \ref{poisson-picture}. \vspace{-0.3cm}}
    \label{poisson-basic-global-numbering}
\end{figure}

We can then create the local numbering for this element by labelling each vertex, edge and cell, and creating a dictionary for the node numbers for each entity. In Figure \ref{lagrange-element-numbered}, the topological entities are already named, so the numbering is the one given below, where the keys of the dictionary are the number of the topological entity, and the entries are the corresponding nodes.
\vspace{-0.2cm}
\begin{align*}
    &\text{Vertices:} \{0: \{0\}, 1: \{14\}, 2: \{4\} \} \\ 
    &\text{Edges:} \{0: \{5, 9, 12\}, 1: \{1, 2, 3\}, 2: \{8, 11, 13\} \} \\ 
    &\text{Cells:} \{0: \{6, 7, 10\} \} 
\end{align*}
Using the local numbering, a possible global numbering of the mesh in Figure \ref{poisson-picture} can also be created. The simplest way is to start by locally numbering one cell, and then continue the indices while numbering the other. The result of this global numbering is shown in Figure \ref{poisson-basic-global-numbering}. There are multiple ways to implement global numbering, as there are in local numbering, so once again the example in Figure \ref{poisson-basic-global-numbering} is just one method that could be used.

This global numbering looks identical to the local numbering of two elements put together, but for some elements this will not be the case. For example, a vector function space or function spaces for more complex elements may have multiple degrees of freedom at each point in the mesh. 

A solution to a variational problem is calculated on the global scale, using a global vector or matrix derived from the global numbering. This can then be moved back to the local scale, to find the solution on each element. 

Firedrake has an internal system to go between the local and global numbering, to preserve continuity between elements, and more fundamentally to correctly make the solution through the correct identification of where each node is through these translations. This is the concept of a local to global map, defined below \cite[pp. 85]{fenics-book}. 

\begin{definition}
    Local to Global Map: A map from local degrees of freedom to global degrees of freedom, specifying how the cells are put together. 
\end{definition}
For example, continuing with the Poisson example of Section \ref{section-fem} using the Lagrange of degree 4 example on the unit square with 2 cells, the following indices are given for the local to global map when considering the Dirichlet boundary conditions on the left and right-hand side of the unit square:
\par
\centerline{\texttt{[-1 1 2 3 -1 -1 6 7 8 -1 10 11 -1 0 -1 -1 16 17 18 -1 20 21 -1 23 -1]}}
\par 

The indices with $-1$ in correspond to the degrees of freedom belonging to nodes on which a Dirichlet boundary condition is applied, and the other indices in the global numbering get mapped to themselves. The indices containing $-1$ will result in the effect of when a matrix is assembled, calculations on that row will not update the values on the row as these rows are skipped over, but these indices still contribute to the size of the matrix.

\subsection{Introduction to PETSc: Sections and DMPlex}\label{section-petsc}
PETSc \cite{petsc-web-page} is a large library which is used to solve multiple types of mathematical problems. It is especially relevant in the case of the report for the use of PETSc in the solution of partial differential equations, and for solving these equations in parallel, using MPI \cite{PETSc-Numerical-Book}. Firedrake uses PETSc, through the petsc4py Python library \cite{petsc4py}, to create various structures for use in providing data to different classes. The main structures that are of interest in this section and report are \texttt{Section}, \texttt{DM} and \texttt{DMPlex}. \par

The \texttt{Section} object, in the mode that it is used in this report, is responsible for creating mappings from points to degrees of freedom associated with these points \cite[pp. 203]{petsc-user-ref}.
A point, in PETSc terms, is an integer representing an index \cite[pp. 203]{petsc-user-ref}. In the context of solving a problem using Firedrake, the point will be one of the topological entities on the mesh. The offset of a mesh point, which is the other piece of information given in the \texttt{Section}, is the index of the first degree of freedom on that entity. Additionally, the information in the \texttt{Section} contains the dimension of each point, which specifies the number of degrees of freedom associated with that point. This means that it is possible to ascertain what points and degrees of freedom are associated with each other, but given just the \texttt{Section}, there is still no information about the connections between mesh points. To be able to know which vertices connect to which edges, another structure defined on the mesh has to be examined, which is the PETSc \texttt{DMPlex} object. \par

The \texttt{DM} structure is an overarching structure which supports multiple types of data model \cite[pp. 201]{petsc-user-ref}. The one that Firedrake uses most frequently is \texttt{DMPlex}, one of the multiple subclasses of the \texttt{DM} structure. In the documentation for PETSc, the brief description of \texttt{DMPlex} is that it ``allows the user to handle an unstructured grid" \cite[pp. 202]{petsc-user-ref}. The advantage of this is that programming PDE solvers can become ``dimension independent", as the treatment of the topological entities is the same regardless of the dimension \cite[pp. 202]{petsc-user-ref}. This allows the use of the same algorithm for different meshes \cite[pp. 93]{matt-knepley-book}, crucial for development of generic libraries over more specific mesh-related code. \par

A \texttt{DMPlex} structure is used in Firedrake to record the relations between the mesh points given in the \texttt{Section}. Through the report, we will shorten \texttt{DMPlex} to \texttt{DM}, as is the convention in the code. Through looking at this particular \texttt{DM}, it is possible to reconstruct the mesh with the correct orientation. 
\begin{figure}[h!]
    \centering
    \includegraphics[scale=0.33, trim={3.9cm, 2.8cm, 3.9cm, 3.5cm}, clip]{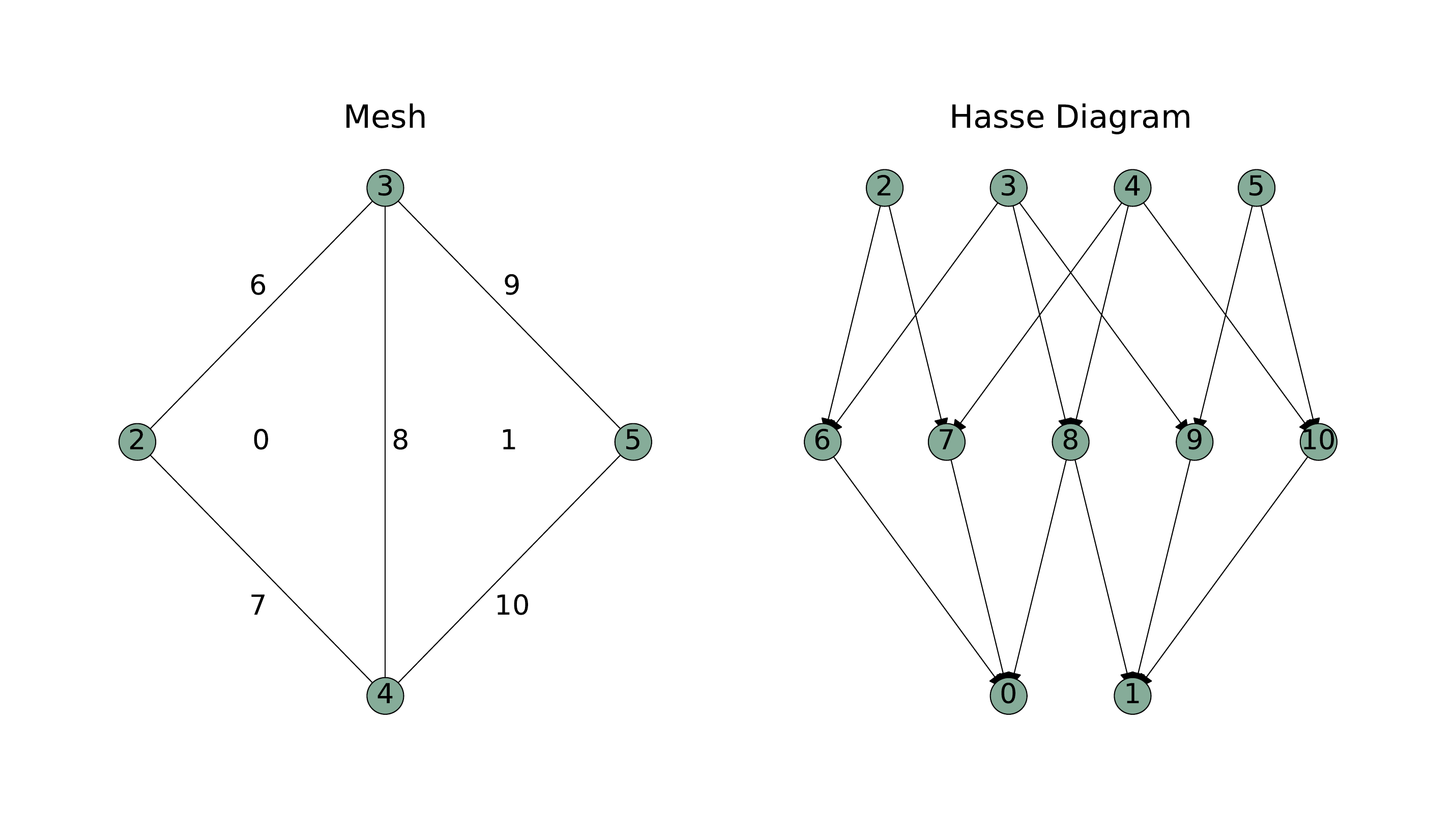}
    \caption{A Hasse diagram (right) for a simple doublet mesh (left). The top layer of the Hasse diagram represent the vertices, the middle layer the edges and the bottom layer the cells. \vspace{-0.5cm}}
    \label{doublet-mesh-and-hasse-diagram}
\end{figure}

To represent such an unstructured mesh graphically rather than in code with a \texttt{DM}, we introduce Hasse diagrams. These are Directed Acyclic Graphs, with edges representing covering relations between the points of the graph \cite[pp. 94]{matt-knepley-book}. The graphs are a useful way to show unstructured meshes, and to visualise the operations that can be performed on their corresponding \texttt{DMPlex} structure responsible for representing the mesh. A Hasse diagram, in the context of a 2D mesh, will have 3 ``layers", one for each dimension of topological entities (vertices, edges, and faces). If the mesh was in a higher dimension, there would be more layers following analogously. Figure \ref{doublet-mesh-and-hasse-diagram} reproduces the 2D doublet mesh and Hasse diagram in the PETSc manual \cite[pp. 201]{petsc-user-ref}. \par

\begin{figure}[h!]
    \centering
    \includegraphics[trim={3cm, 2cm, 1.5cm, 2.2cm}, clip, scale=0.4]{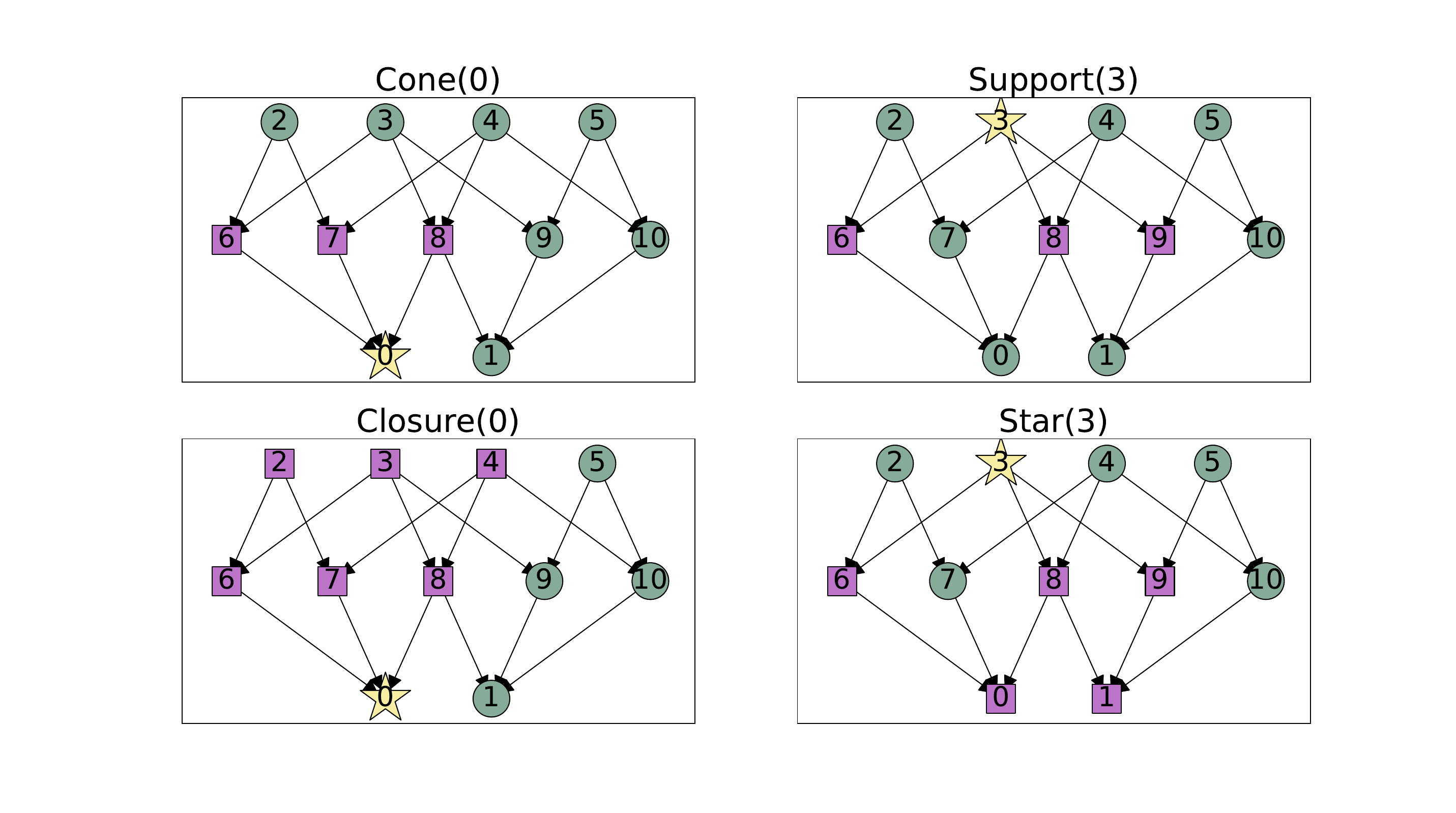}
    \caption{Four Hasse diagrams, the same as Figure \ref{doublet-mesh-and-hasse-diagram}, representing various common DMPlex operations. The yellow star is the input point for the operation and the purple squares are the output points. The operation name is shown in the title of each subplot.}
    \label{hasse-diagram-operations}
\end{figure}

There are multiple operations that will return information from the \texttt{DMPlex} object, which can assist in identifying the relations between mesh points (\texttt{p}). For example, in this section, the following operations are shown, the definitions that are given are the same as described in \cite[Section 7.1-7.1.1]{matt-knepley-book}. The results of the following operations are shown in Figure \ref{hasse-diagram-operations}.
\begin{itemize}
    \vspace{-0.3cm}
    \item \texttt{DMPlexGetCone(p)}: Returns the set of points that cover \texttt{p}. The points that cover \texttt{p} are in the next dimension lower than the dimension of \texttt{p}.
       \vspace{-0.3cm}
    \item \texttt{DMPlexGetSupport(p)}: Returns the set of points that are related to (are covered by) \texttt{p}. The points that are covered by \texttt{p} are in the next dimension higher than the dimension of \texttt{p}. 
       \vspace{-0.3cm}
    \item \texttt{DMPlexGetClosure(p)}: The transitive closure of cone, meaning this returns all points that cover \texttt{p}, and then all the points that cover those points recursively until the lowest dimension.
    \vspace{-0.8cm}
    \item \texttt{DMPlexGetStar(p)}: The transitive closure of support, meaning this returns all points that are covered by \texttt{p}, and then all points that are covered by those points recursively until the highest dimension.
    \vspace{-0.4cm}
\end{itemize}

\begin{figure}[h!]
    \centering
    \includegraphics[trim={6.7cm, 3.6cm, 5.5cm, 2cm}, clip, scale=0.35]{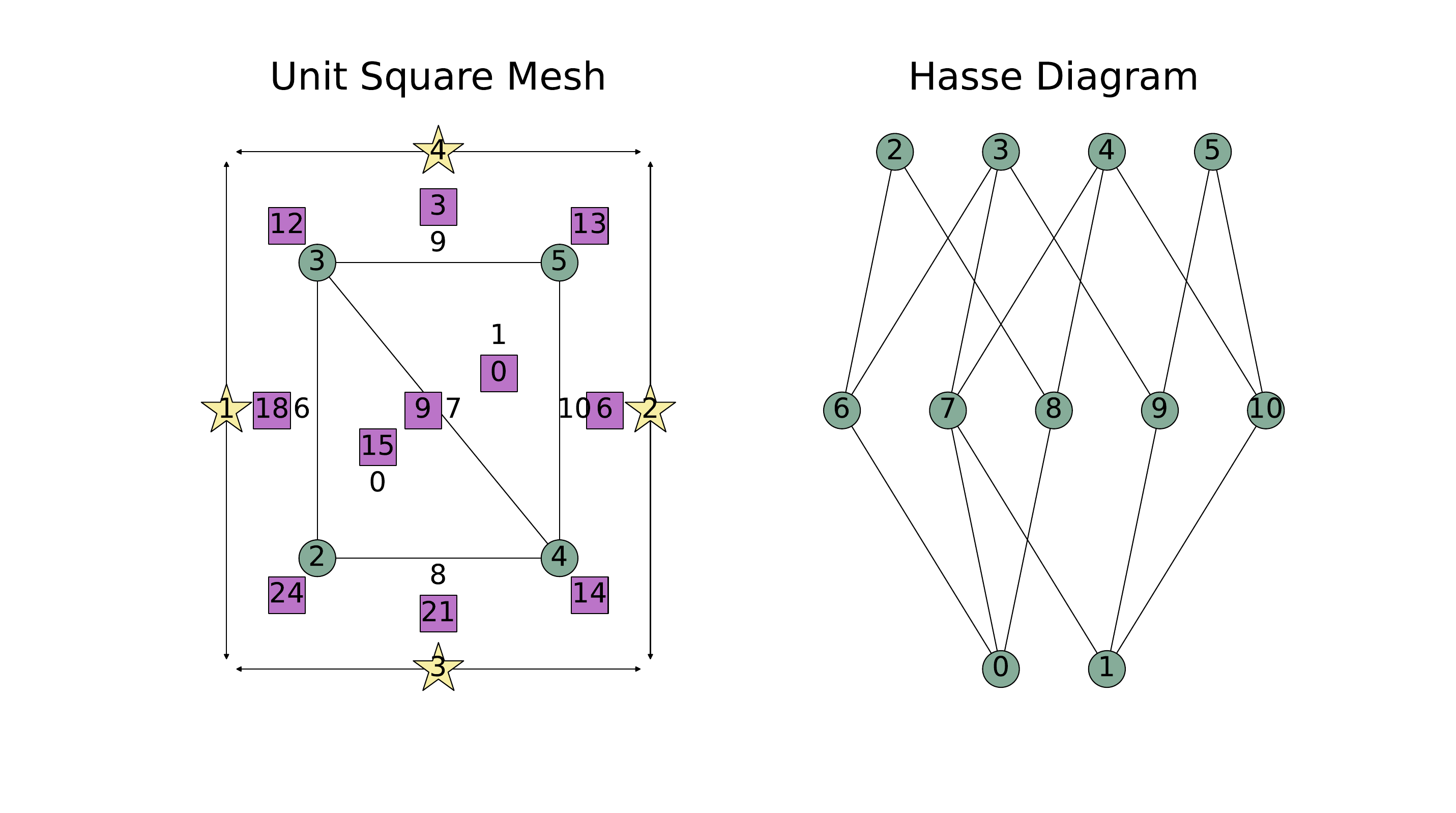}
    \caption{Left: Numbering of the mesh made of 2 Lagrange elements of degree 4, as seen in Figure \ref{poisson-picture}. The purple squares represent the offset at each point, the yellow stars represent the subdomain of the boundary mesh. Right: Corresponding Hasse diagram for the 11 mesh points.}
    \label{unit-square-mesh-hasse-diagram}
\end{figure}

Additionally, labels are added to each mesh point, providing even more information about the point. For example, Firedrake adds a ``Face Sets" label, which groups mesh points according to their subdomain, allowing for the correct orientation of the mesh when using a mesh where the numbering of the subdomain is known.  

To see how the \texttt{DM} and global numbering objects are linked, a recreation of the numbering for the fourth-degree Lagrange element on the mesh is seen in Figure \ref{unit-square-mesh-hasse-diagram}. The diagram in Figure \ref{unit-square-mesh-hasse-diagram} shows both the points in the \texttt{Section} sees them, and the offsets, with information from the \texttt{DM} to make sure the square is the correct orientation and points are connected in the right way. 


In Figure \ref{unit-square-mesh-hasse-diagram}, it is not possible to determine which degree of freedom on an edge corresponds to the exact offset. For example, the only information obtained is that point 6, an edge, has the degrees of freedom 18, 19 and 20. This could be 18, 19, 20 going from top to bottom or vice versa, but there is no way to definitively state which orientation is the correct one.
The correct ordering of these degrees of freedom can be found by interpolating $x$ or $y$ into the mesh. This is not relevant to the rest of the report, so the results of this operation are omitted for brevity.

To summarise, below is a list containing the two main structures discussed in this section that will be featured in the report: 
\begin{itemize}
    \item Global Numbering: A PETSc \texttt{Section} object mapping mesh points to degrees of freedom, with the degrees of freedom coming from the function space. This is created in the \texttt{create\_section} routine.
    \item DM: A PETSc \texttt{DMPlex} object containing the details on the structure of the mesh. This is created as the mesh is created, and is stored as an attribute of the mesh. 
\end{itemize}

This focus on the numbering of a mesh seems circular, given that this chapter is dedicated to effectively reconstructing a mesh that we know the orientation of already, which is not necessary to solve a problem. However, the link between the offsets, mesh points and other labels assigned to the mesh points are to be brought up again as a change in the implementation section of the report, so it is essential that the reader can understand the different labels available to a mesh point, and what information they provide.

\subsection{PyOP2 and Parallel Communication}\label{section-pyop2}
The mesh points defined in the PETSc \texttt{Section} have one more useful label applied to them in the \texttt{DM}, which is the label of their PyOP2 type of point. 
The purpose of this labelling is to ensure the correct transfer of data between processes in parallel. A computation in parallel happens when more than one process is working on a problem. 

For example, Firedrake splits up the mesh across the correct amount of processors, when called in parallel, with each process performing calculations on a certain section of the mesh. This done automatically through the use of \texttt{DMPlex}, using a star forest object from PETSc \cite[pp. 5-6]{mesh-management-paper}, which will be revisited later. This is used to share data between processes in a process called a halo data exchange. This data has to be shared rather than stored separately, as adjacent degrees of freedom or mesh points may influence calculations. This creates the distinction between ``core" and ``owned" mesh points. Calculations can still occur in the ``core" points while waiting for the termination of a data exchange that may affect the ``owned" points \cite[pp. 7]{mesh-management-paper}. In addition to ``core" and ``owned", the website for the documentation for PyOP2 lists four distinct types of mesh point \cite{pyop2-mpi}:
\begin{itemize}
    \vspace{-0.25cm}
    \item Core
        \vspace{-0.25cm}
    \item Owned
        \vspace{-0.25cm}
    \item Exec Halo
        \vspace{-0.25cm}
    \item Non-Exec Halo
\end{itemize}
\vspace{-0.25cm}
However in practice, when creating PyOP2 objects such as sets, the latter two labels get condensed into a ``ghost" label. For the rest of the report, the Exec/Non-Exec points will be referred to by ``ghost" points, so there are only three labels to consider in keeping with the labels used by Firedrake. 

\begin{figure}[h!]
    \centering
    \includegraphics[trim={3.5cm, 1.75cm, 3.8cm, 2cm}, clip, scale=0.3]{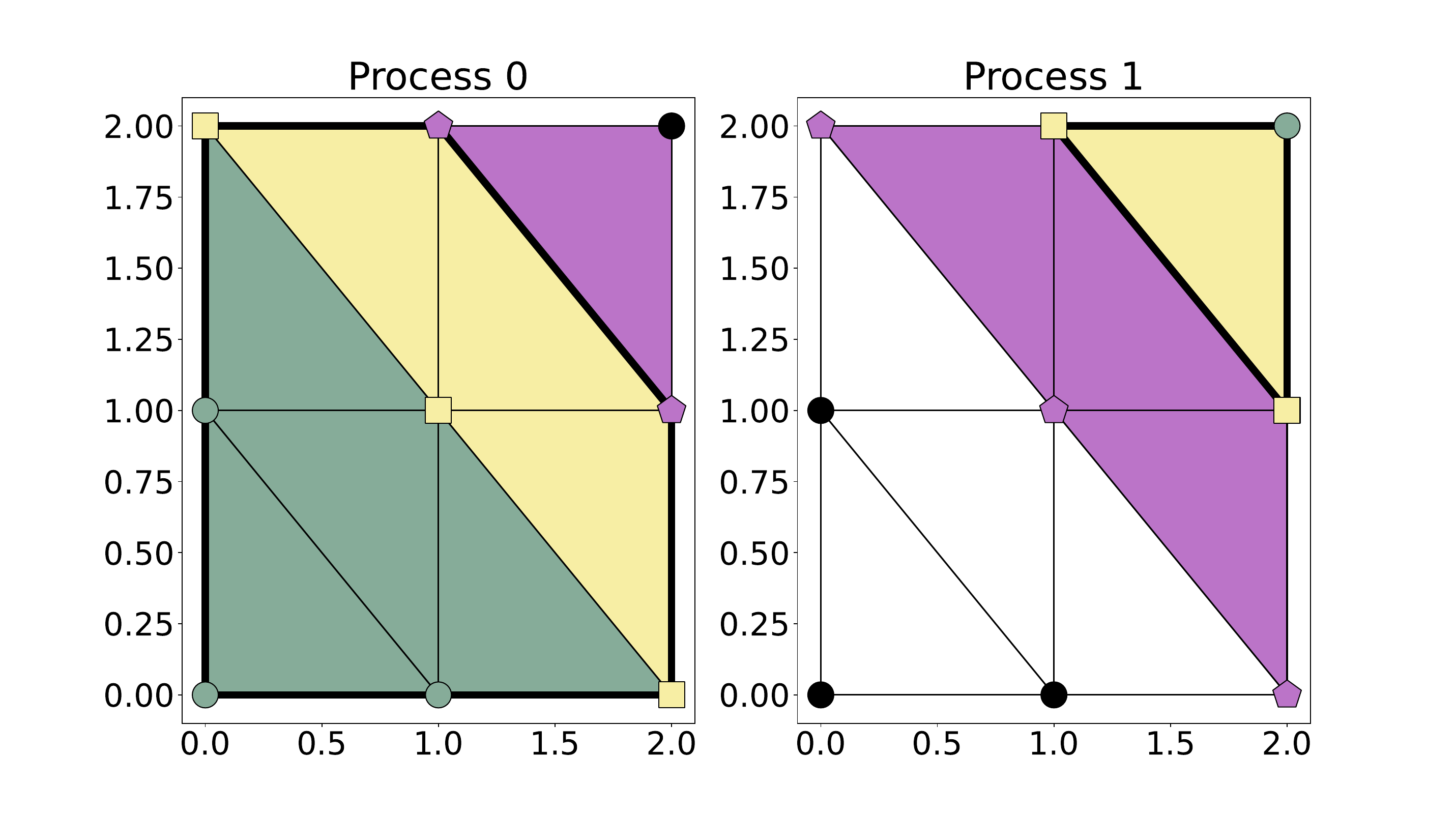}
    \caption{A potential split of a $2\times2$ mesh, on $[0,2] \times [0, 2]$ over two processes in parallel, using degree 1 Lagrange elements. Green circles represent the core points, yellow squares represents owned points and purple pentagons represents ghost points. Cells are highlighted similarly for convenience. The thick line highlights the cells that the process owns, and white cells/black nodes are not used within that process.}
    \label{pyop2-labelling-cg4-unrestricted-2-processes}
\end{figure}

As an example, Figure \ref{pyop2-labelling-cg4-unrestricted-2-processes} is a diagram showing the PyOP2 labelling of a square mesh consisting of 4 cells from, when using two processes in parallel. As shown in the diagram, the ghost areas in one process are the core or owned areas in the other. \par
If the point on the left of the top row on Process 0 in Figure \ref{pyop2-labelling-cg4-unrestricted-2-processes} changed in value, this would update the ghost value of the corresponding point on Process 1 when a data exchange occurs. Evidently, transfer the other way from core/owned points on Process 1 to ghost points on Process 0 works as well. It is important to note that the mesh points do not take the same values in the two processes, as seen in the figure. This means there is the need to be aware of potentially multiple labels used on each point in the future and how they interact, but this will be handled in the corresponding parallel section in the implementation chapter. 

\begin{figure}[h!]
    \centering
    \includegraphics[trim={6.6cm, 4cm, 5cm, 3.5cm}, clip, scale=0.35]{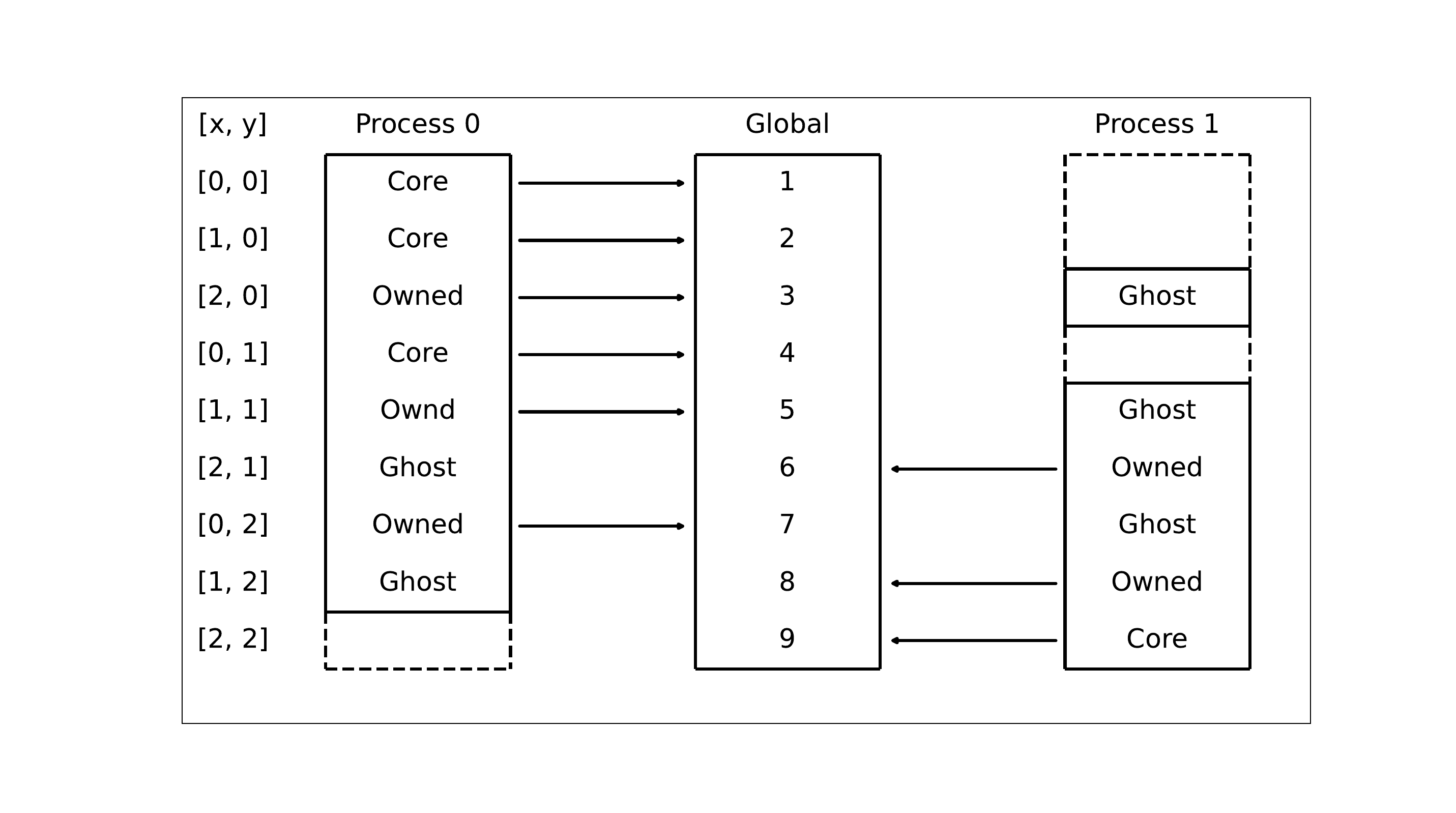}
    \caption{An example of how the points in Figure \ref{pyop2-labelling-cg4-unrestricted-2-processes} (left, right) could combine to give a global numbering (in the centre). The dashed lines represent points which are not represented in that process.}
    \label{parallel-mapping-diagram}
\end{figure}

Additionally, the local to global resulting from a process split in parallel will contain only the core and owned points of each process, meaning that any ghost points will not be found in the global vector. This results in a global vector that is smaller than the two local vectors put together, as the local vectors include the ghost points. Figure \ref{parallel-mapping-diagram} shows how the two processes of Figure \ref{pyop2-labelling-cg4-unrestricted-2-processes} combine to give the global numbering.

PyOP2 is also used to improve the efficiency of calculations in Firedrake elsewhere. This idea will be explored further in detail in Section \ref{section-pyop2-changes}, when objects from PyOP2 are used. 

\subsection{Function Spaces in Firedrake}\label{section-function-spaces-in-firedrake}
When solving a PDE using the finite element method, there will need to be a function space that the solution is in. The role of the \texttt{FunctionSpace} class in Firedrake is to provide this space, as well as the space for the test and trial functions used \cite[pp. 3]{FiredrakeUserManual}.

In Firedrake, there are multiple types of function spaces implemented, with some being scalar or vector only and others being for specific use cases. The main class, \texttt{FunctionSpace}, is constructed using a mesh and a finite element to allow Firedrake to create and manipulate data required to solve a problem. This puts the definition closer to that of the finite element space than a pure function space. 

In the \texttt{FunctionSpace} itself, there are only a few attributes stored, apart from objects such as the specific \texttt{DM} for the \texttt{FunctionSpace} and the equivalent UFL function space. One piece of information stored directly on the function space is the local to global map, which is the mapping described in Section \ref{section-local-global}. Given a list of boundary conditions, the correct local to global map is returned for the \texttt{FunctionSpace}, with the value of the global index set to $-1$ for nodes on the boundary where boundary conditions are applied, which is as expected. 

A class that is linked to a \texttt{FunctionSpace} object is the corresponding \texttt{FunctionSpaceData} object, created within the constructor for \texttt{FunctionSpace}. This object stores much of the data that will be important in the following sections. One such piece of data is the \texttt{global\_numbering} attribute for \texttt{FunctionSpaceData}. This is created, at its base level, by a function called \texttt{create\_section}. The created section describes the global numbering on the given mesh through a PETSc \texttt{Section} object storing the mesh points and the corresponding offsets. 
\par
A significant amount of the data that is created and stored in the \texttt{FunctionSpaceData} object is cached using various cache keys. 
The content of the keys, as in the previous versions of Firedrake, is not of great importance to this report apart from the general idea that they are based on attributes that aim to be sufficient to distinguish between function spaces. This explains why so much of the data the \texttt{FunctionSpace} uses is stored in the \texttt{FunctionSpaceData}: it is so that 2 instances of the same \texttt{FunctionSpace} can access the same data, allowing the creation of data to only be done once. With this principle in mind, only the data that is required to mark out a specific \texttt{FunctionSpace} such as the name of the space, or objects that do not necessarily need to be cached such as the UFL element, are stored on the \texttt{FunctionSpace} itself. 
This key does not currently contain information on what boundary conditions have been or will be applied to functions defined on the \texttt{FunctionSpace} object, as for now the numbering is not affected. This will be revisited in Section \ref{section-implementation}. 

As well as the \texttt{FunctionSpaceData} object, the mesh itself has a role in storing data that the function space may need to access, as has been seen in Section \ref{section-petsc} with the DM. This is important to remember through the implementation process, as it means that the modification of data structures used in the \texttt{FunctionSpace} object has consequences throughout the code base that need to be taken into account. 

\section{Implementation of RestrictedFunctionSpace in Firedrake}\label{section-implementation}
\subsection{Introduction}\label{section-impl-intro}
To solve the issues described in Section \ref{section-motivation}, it is useful to review the splitting of the function space based on Dirichlet boundary conditions. We have already seen in Section \ref{section-firedrake-bcs} that given a Dirichlet boundary condition on a subdomain $\Gamma$ we can split the function space $V$ into $V_{0}$ and $V_{\Gamma}$. As described before, $V_{0}$ is spanned by all basis functions which vanish on $\Gamma$ and $V_{\Gamma}$ is spanned by all other basis functions. This means all nodes and degrees of freedom corresponding to $\Gamma$ are in $V_{\Gamma}$ and so correspond to the boundary nodes through the definition of the nodal basis.

From this decomposition, we can see that if a function space of $V_{res} = V_0$ is created when assembling a matrix or solving a problem, the following equation from Equation \ref{final-update-newton-J-decomposition} is returned:
\begin{equation*}
    J^{00}\hat{U} = F(u)
\end{equation*}
The terms involving $\Gamma$ disappear due to there being no $V_\Gamma$ when using $V_{res}$. As an immediate benefit, this update step is easier to read than the original equations. It also allows only the unknown values to be solved for, rather than all node values as previously required, and removes the slight deception present in the code of ignoring initial boundary values the user supplies for $u$.

Similarly, when looking at these problems with the eigensolver, it is clear that again if we were to set $V_{res} = V_{0}$ for our arguments in the form, the issue would be resolved. This is because if the boundary degrees of freedom did not appear in the assembled matrix, we would have that the boundary eigenvalues are not present to begin with and clashes would no longer be an issue. The user could be confident that the eigenvalues they get as output are not ones that have come from a shift of the eigenvalues associated with the boundary. 

These two, identical, solutions to the problems raised suggests that a new object should be made that is similar to the function space class currently in Firedrake. However, in taking trial and test functions for the variational problem or eigenproblem, the space these functions come from is $V_0$ rather than $V$. The output solution should still be in $V$ when solving a variational problem, so that the boundary conditions are applied and respected in the solution. Even more explicitly, in the restricted space, we assemble the following forms for a linear variational problem:
\begin{equation*}
    a(u_h, v), G(v)  \text{ where }u_h, v \in V_0 
\end{equation*}
However, the output of \texttt{solve} will return $u_h$ as a function in $V$, through adding in the rest of the nodes at the end. 
By the decomposition of our original function space $V$, this $V_0$ space is achieved by removing only the boundary degrees of freedom from consideration when solving problems.

The mathematical description of this solution means that in code, it is required to mark out these boundary degrees of freedom in order to effectively remove them. As a consequence, the majority of the changes made to Firedrake when implementing this new class are tying boundary markers, identifying which regions of the mesh will contain Dirichlet boundary conditions, to the function space itself. 
An extra requirement is that users can continue to use the same methods of implementing boundary conditions that existed in Firedrake from before the project, as users may have various dependencies on old code, meaning that the changes should be separate from the already-made \texttt{FunctionSpace} object. For this, \texttt{RestrictedFunctionSpace} is a new class, inheriting from \texttt{FunctionSpace}. Due to the solutions mentioned previously, \texttt{RestrictedFunctionSpace} will be one of two representations depending on where the space is appearing:
\vspace{-0.3cm}
\begin{itemize}
    \item The restricted space should still give the boundary node values when returning a solution to a problem, meaning the output space still behaves like $V$.
    \item The restricted space should only consider trial functions and test functions from $V_{0}$ when assembling and solving a problem. 
\end{itemize}
\vspace{-0.3cm}
The description of how this restricted space will be implemented in Firedrake to match this specification is detailed in the next sections. 

This section will also discuss the features that are now in Firedrake that use the \texttt{RestrictedFunctionSpace} class. These will be detailed at the end of this section, as these are more qualitative and do not add any more details as to how the class was implemented.
 
Now that the basic ideas of what the \texttt{RestrictedFunctionSpace} needs to do have been consolidated, the minimum working case for a \texttt{
RestrictedFunctionSpace} such that the class is better, or equivalent, to use than a regular \texttt{FunctionSpace} should be considered, to give another metric to base the success of the project against. At a minimum, the \texttt{RestrictedFunctionSpace} should be able to be used in solving variational problems when using both homogeneous and inhomogeneous Dirichlet boundary conditions, on most common element types. The main exception to this is the Hermite element on the unit interval, as the way boundary nodes are calculated is different to every other supported element in Firedrake. 

It would be advantageous to be able to multiply restricted spaces together to get a mixed space, similar to how a \texttt{FunctionSpace} can be multiplied. This should also work for a multiplication between restricted and unrestricted spaces. Doing this, as well as adding in an equivalent restricted \texttt{VectorFunctionSpace}, would mean that a \texttt{RestrictedFunctionSpace} does not have less utility than a standard \texttt{FunctionSpace}.

As the \texttt{RestrictedFunctionSpace} class is the response to an issue arising in the eigensolver when using regular \texttt{FunctionSpace} objects, the eigensolver should be changed so that the appropriate restricted space is always used regardless of what spaces are used in the definition of the forms coming into the eigensolver. This should always be done, unless the user specifies otherwise, to limit convergence failure and other frustrations mentioned in the Section \ref{section-motivation}. A similar setup for the \texttt{solve} function will be useful, but this feature can be considered as not as essential as the corresponding feature in the eigensolver.

\subsection{Changes to Firedrake}\label{section-firedrake-changes}
When describing fully how \texttt{RestrictedFunctionSpace} works, there are broadly two sections that need to be discussed: the creation of the class itself and the modification of existing functions that the class depends on to work correctly. 

The first section is the simpler section, and will be the focus of the first subsection. This is where the main choice in how to differentiate the \texttt{RestrictedFunctionSpace} from a \texttt{FunctionSpace} itself is made, and details how the user can create a \texttt{RestrictedFunctionSpace} for themselves. The second section is a lot longer, and where most of the changed algorithms and functions will be discussed. Given the choices made in what is used for the class, we have to then integrate what we want to do with what is currently there. This is where a lot of the PETSc objects mentioned in Section \ref{section-general-implementation} will come in, as we relabel mesh points and offsets to best achieve our goal. The user would not typically see this part, except as attributes of the class. 

\subsubsection{Top-Level Class Implementation}\label{section-top-lvl-firedrake}
The description of \texttt{RestrictedFunctionSpace}, as discussed in Section \ref{section-impl-intro}, means that it was evident that the class should inherit from the generic \texttt{FunctionSpace} class of Firedrake. A \texttt{RestrictedFunctionSpace} is a \texttt{FunctionSpace} with an added piece of data given by the user: a set of markers of the mesh detailing where they intend to apply Dirichlet boundary conditions. Due to this extra information, the first function of \texttt{RestrictedFunctionSpace} that diverges from the superclass is the constructor. In the \texttt{RestrictedFunctionSpace}, the boundary set is marked down as an attribute of the class, called \texttt{boundary\_set}. The rest of this section is almost entirely dedicated to following what needs to change at each layer of Firedrake, and the key functions that produce modified results in the presence of a boundary set. 
\par 

As a \texttt{RestrictedFunctionSpace} object has the \texttt{boundary\_set} attribute, this attribute is also given to the associated \texttt{FunctionSpaceData} object for use in creating the data that the \texttt{RestrictedFunctionSpace} relies on. To achieve this, a \texttt{boundary\_set} parameter is now present in the constructor of \texttt{FunctionSpaceData}, defaulting to \texttt{None}. This can then be easily passed through to further functions, and set as an attribute of \texttt{FunctionSpaceData}, with the same name. This has two useful effects:
\begin{itemize}
    \item As discussed in Section \ref{section-function-spaces-in-firedrake}, many of the objects stored inside the \texttt{FunctionSpaceData} are cached, with the key depending on element and mesh specific attributes. To return the \texttt{Section} associated with a \texttt{FunctionSpace} the key is now \texttt{(nodes\_per\_entity, real\_tensorproduct, boundary\_set)}.
    The addition of the boundary set to the cache key allows for the differentiation between a \texttt{FunctionSpace} and a \texttt{RestrictedFunctionSpace} on the same mesh and element. This is important in making sure the correct attributes are returned for the correct space. Through the rest of the implementation section, it will become clear that some key attributes of the \texttt{FunctionSpaceData}, such as the global numbering, changes depending on if the data is created on a restricted or unrestricted space.
    \item These objects which are created inside the \texttt{FunctionSpaceData} may require the location of the Dirichlet boundary conditions passed into them, in order to use them in the calculations made with respect to numbering the mesh. The passing down of the boundary set into the \texttt{FunctionSpaceData} allows these functions to access this value easily. 
\end{itemize}

The next sections will look at how this \texttt{boundary\_set} is used throughout Firedrake to obtain the correct relabelling. 

\subsubsection{PETSc and Cython Layer: Relabelling Mesh Points and Offsets}\label{section-cython}
A solution to the issue of the treatment of Dirichlet boundary conditions in Firedrake has already been considered in other work, \cite[pp. 71-72]{matt-knepley-book}, which involves the removal of boundary values from the global vector where the problem is being solved. The suggestion code-wise from this book to avoid solving for boundary values at the global level is to mark the boundary values as constrained, which is a PETSc label. This solution is the same as the solution as the ideas presented in Section \ref{section-impl-intro}, so in the code this idea is what the \texttt{RestrictedFunctionSpace} class implements at the PETSc level. This section will focus on how this is done in Firedrake for the \texttt{RestrictedFunctionSpace} class, and what marking a node as constrained means.

As there is a boundary set used in the definition of a \texttt{RestrictedFunctionSpace}, the functions that create the labelling of the mesh, as seen in Sections \ref{section-petsc} and \ref{section-pyop2}, require the boundary set in their definition as well. This is to mark the mesh, to be able to exclude the boundary degrees of freedom from calculations while including the necessary ones. This section will refer to boundary degrees of freedom as being constrained, or sometimes as constrained nodes, due to the terminology that PETSc itself uses. In some cases not all degrees of freedom on a boundary node are constrained, such as in Hermite elements or vector-valued elements, but in our Lagrange element example the definitions are effectively interchangeable. 

From Section \ref{section-pyop2}, it is known that each mesh point has a label; one of ``core", ``owned" and ``ghost". The number of core or owned points sets the size of the matrix created through \texttt{assemble}, as those points are the ones the process owns and has control over. In the function \texttt{plex\_renumbering}, Firedrake orders mesh points based on these PyOP2 labels. ``Core" points get moved to the front, then followed by the ``owned" points and finally ``ghost" points to the end. 
The goal of the class, as stated in the motivation section, is to be able to take out the constrained degrees of freedom, to reduce the size of the basis of the function space. This leads to the conclusion that boundary degrees of freedom need to be pushed to the end of the ``core" and ``owned" block, so that they can be easily identified, counted and cut off. As \texttt{plex\_renumbering} sorts based on the PyOP2 labels of the mesh points, the movement of any constrained points occurs here as well. This means that an extra key is added to the sort, to determine if the ``core" or ``owned" point is also restricted. The \texttt{plex\_renumbering} algorithm is an example of a lexicographic sort. In the presence of boundary degrees of freedom, the ordering of the sort of the indices based on both the PyOP2 label and if the point is on the boundary is the following:
\begin{equation}
 \text{core, not boundary} <  \text{owned, not boundary} <  \text{core/owned, boundary}  < \text{ghost}
\end{equation}
The ordering is well-defined as all mesh points fall into one of the categories mentioned above. 

\begin{figure}
    \centering
    \includegraphics[trim={0.5cm, 4cm, 0.5cm, 0.5cm}, clip, scale=0.5]{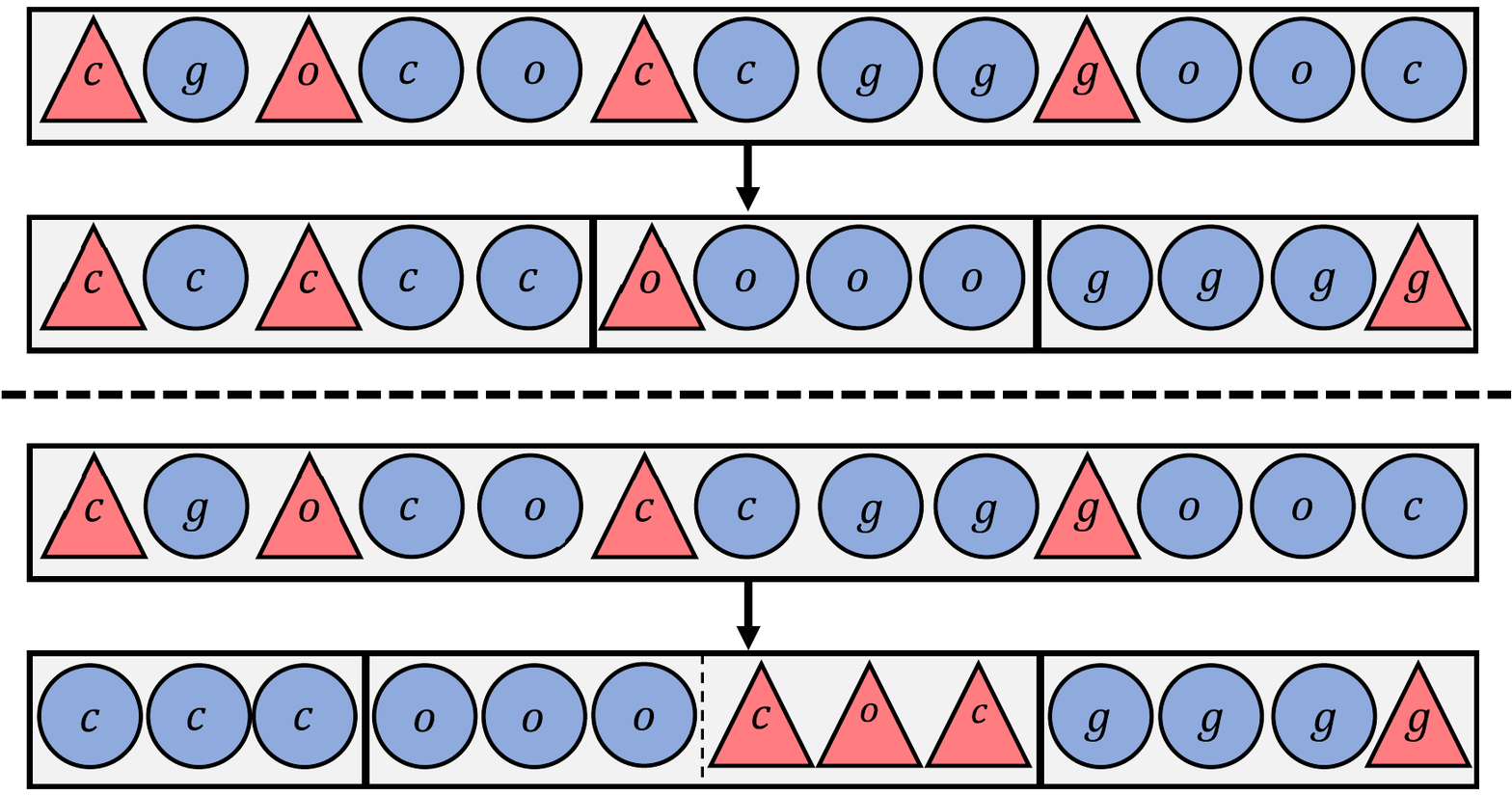}
    \caption{Top two rows: Plex renumbering on a mesh with a regular function space, assuming left-to-right ordering. The input is the top row and the output is the second row. Bottom two rows: Plex renumbering on a mesh with a restricted function space. All core and owned boundary points (red triangle) are now pushed to the back of the core + owned block. The letters correspond to the PyOP2 label of that mesh point.}
    \label{plex-renumbering-diagram}
\end{figure}

A diagram showing how the renumbering occurs for a \texttt{FunctionSpace} object is shown at the top of Figure \ref{plex-renumbering-diagram}. A diagram of the renumbering of the same mesh using a  \texttt{RestrictedFunctionSpace} is in the bottom of Figure \ref{plex-renumbering-diagram}. It should be noted that constrained ghost points do not generally get reordered to the back of the ghost block, but in the case in the diagram this occurs due to the original ordering of the points. 

In the \texttt{plex\_renumbering function}, the start and end indices of the constrained block of core and owned points is calculated and returned, to be used by the \texttt{create\_section} function. In the bottom of Figure \ref{plex-renumbering-diagram}, the starting index is 6 and the ending index is 8 for the constrained block, whereas for the top diagram the two indices are 8 and 8, the same index as there is no constrained block. 

This \texttt{plex\_renumbering} function is used within \texttt{create\_section}, which is responsible for creating the local section. As discussed in Section \ref{section-petsc}, this local section is used to show the mesh points and the corresponding offsets, labelling the index of the first degree of freedom on this mesh point. The ordering of these offsets is based on the ordering of the mesh points provided by \texttt{plex\_renumbering}, meaning that the boundary points should have higher offsets. 

The \texttt{create\_section} function is initially called from inside the constructor of the \texttt{FunctionSpaceData} object associated with the \texttt{FunctionSpace}, and is stored as the \texttt{global\_numbering} attribute.

The differences in the \texttt{Section} created from a function space $V$ and one from a \texttt{RestrictedFunctionSpace} restricting $V$ were that the restricted section contains the reordering of the degrees of freedom on the mesh points as given from \texttt{plex\_renumbering}, and is required to mark points on the boundary as constrained. Documentation for PETSc describes a constrained degree of freedom as one that appears in the local vector but not in the global vector \cite[pp. 205]{petsc-user-ref}, and that has a fixed value. It is a good place to reiterate that in Firedrake, when assemble is called with boundary conditions, all that is different is that the rows for the boundary degrees of freedom are transformed into an identity row. The value of the boundary condition does not matter for this algorithm, and so when assembling a form marking the degrees of freedom as constrained only gives the code another label to use and condition on. However, this marking will matter when solving a problem, as it preserves the value of the solution on the boundary degrees of freedom. In an unrestricted problem, Firedrake already does this, through the use of $F_\Gamma(u; v_\Gamma)$. Due to the renumbering given, the offsets on the boundary points should be greater than the offsets not on the boundary points. 

\begin{figure}[h!]
    \centering
    \includegraphics[trim={3.5cm, 3cm, 2cm, 3cm}, clip, scale=0.4]{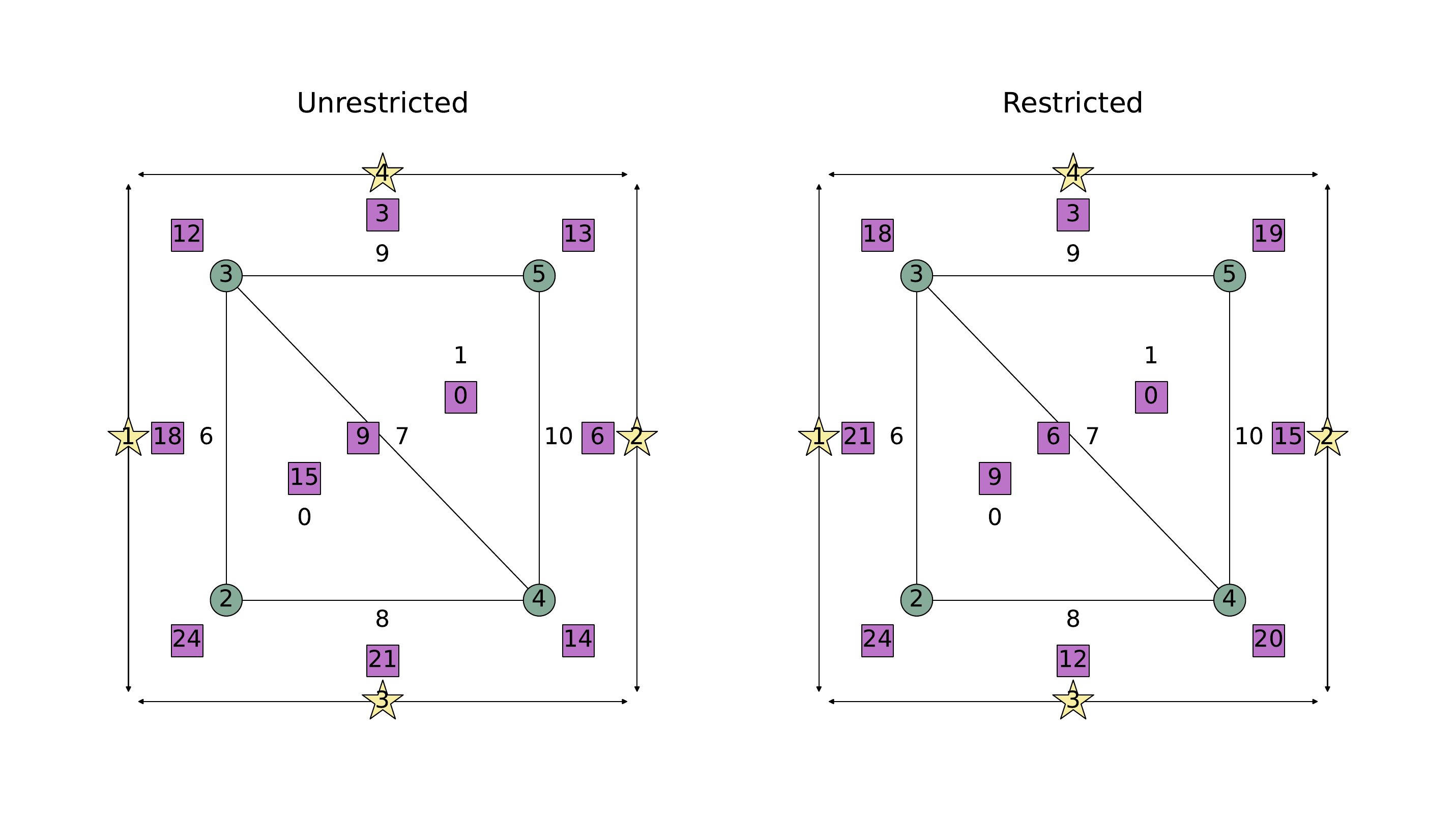}
    \caption{Diagrams of the mesh points (black/circled) and their offsets (purple squares) for a 2-triangle square mesh using fourth-order Lagrange elements. On the left is the numbering for a \texttt{FunctionSpace}, while on the right is the numbering for a \texttt{RestrictedFunctionSpace} with a boundary set of [1, 2] (subdomains indicated by stars). The left figure is the same as Figure \ref{unrestricted-restricted-mesh-diagram}. \vspace{-0.3cm}}
    \label{unrestricted-restricted-mesh-diagram}
\end{figure}

To look at the information given by the \texttt{Section}, Figure \ref{unrestricted-restricted-mesh-diagram} is a diagram of a unit square mesh with 1 element in each direction, using fourth degree Lagrange elements with the different offsets and mesh point numbers labelled in both the corresponding restricted and unrestricted function spaces, similar to Figure \ref{unit-square-mesh-hasse-diagram}. This is created using the serial case, and may depend on any mesh reordering defined. On the restricted diagram, we set the lines $x = 0$ and $x = 1$, or subdomains 1 and 2 of the mesh, to have a homogeneous Dirichlet boundary condition, to mirror the example given in previous sections.

It is important to see that through the creation of the \texttt{Section} by \texttt{create\_section} the layout of the mesh points has not changed. This is because this information is stored on the mesh, through the \texttt{DM} object, rather than the \texttt{FunctionSpace} object itself. However, the numbering of the offsets throughout the section has changed. Looking at the right diagram in Figure \ref{unrestricted-restricted-mesh-diagram}, we can see that all the offsets on the subdomains which contain a boundary condition, $x = 0$ and $x = 1$, have offsets that are at least 15. The conclusion is that these degrees of freedom are now numbered from $15 - 24$, which is what was wanted, as the first $15$ are therefore unconstrained degrees of freedom and are the ones which need to be kept. 

There are some other distinctions between the restricted and unrestricted spaces, such as the string representation, but these are mostly cosmetic and unnecessary to discuss. One other major difference between the two classes of the class itself is the local to global numbering map. The map, as discussed in Section \ref{section-local-global}, contains a set of indices with a -1 in the entries corresponding to boundary nodes. 

\begin{figure}%
    \centering
    \subfloat[\centering Unrestricted]{{\includegraphics[trim={1cm, 0cm, 17cm, 3cm}, width=5.025cm]{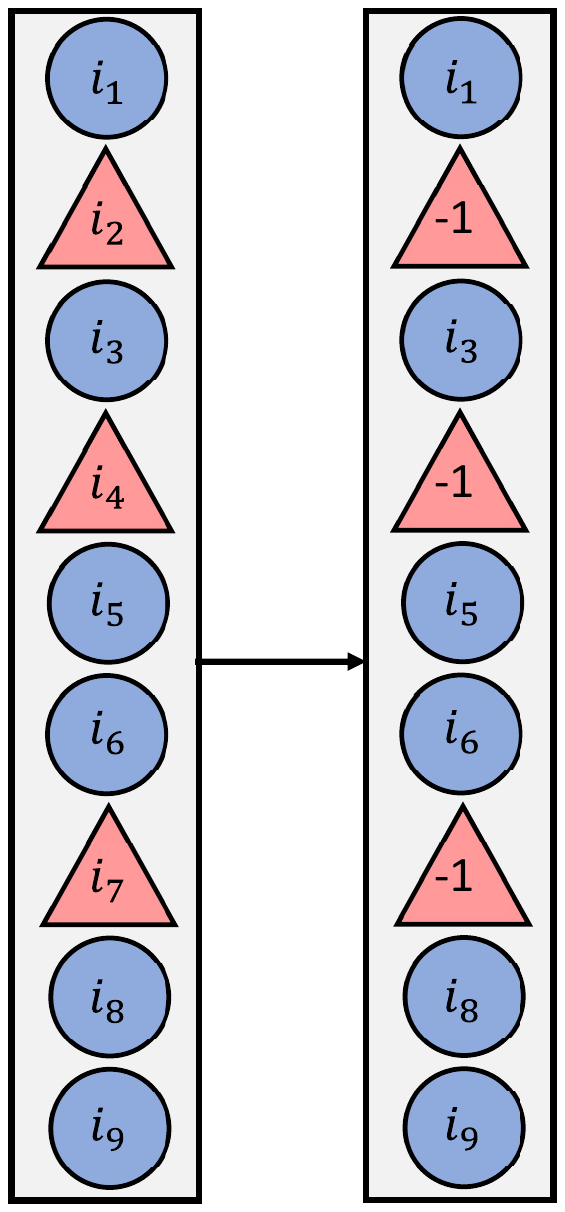} }}%
    \qquad
    \subfloat[\centering Restricted]{{\includegraphics[trim={2cm, 1cm, 18cm, 1cm}, width=3.65cm]{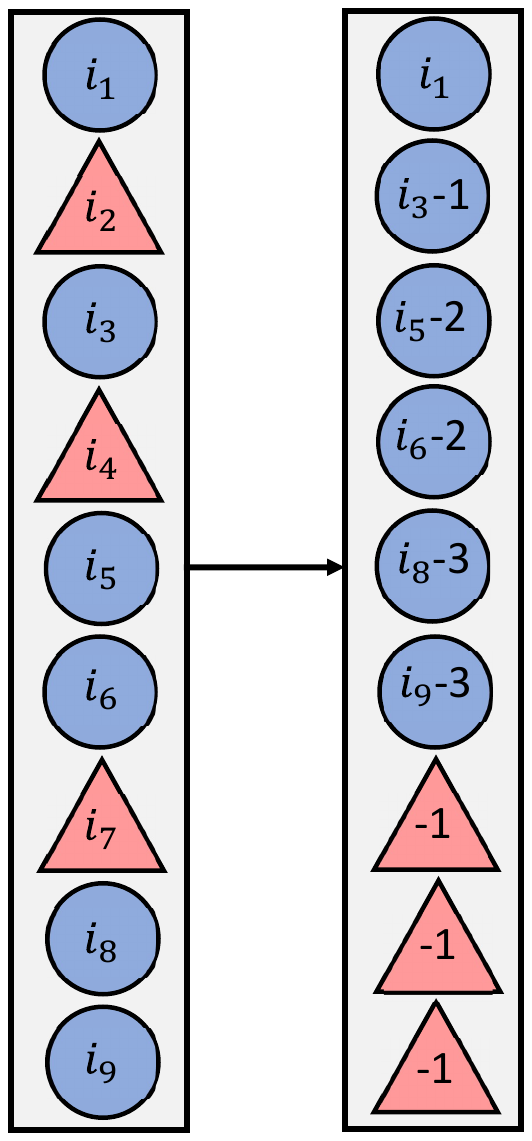} }}%
    \caption{The indices of a local to global map for a (a) \texttt{FunctionSpace}, (b) \texttt{RestrictedFunctionSpace} object. Both diagrams take in the indices on the left, and return the indices on the right. Red triangles are restricted/boundary points, while blue circles are unrestricted points. \vspace{-0.5cm}}%
    \label{lgmaps-figure}%
\end{figure}

Figure \ref{lgmaps-figure} show the difference in the \texttt{LGMap} returned when using a restricted or an unrestricted space. In the restricted case, as all the offsets for the boundary mesh points are now at the end, any index with $-1$ in must be at the end of the indices for the local to global map as well. Additionally, all the other indices must be consecutive, as the degrees of freedom of the unconstrained points are set so that these are the first degrees of freedom. 

The creation of the local to global numbering map for a restricted space is not done in a function at the top-level, as is the case for \texttt{FunctionSpace}. Instead, we use the local to global numbering associated with the halo, even when not running in parallel, where a function called make\_global\_numbering is called from there. We describe this routine in Algorithm \ref{updated-make-global-numbering}.

\begin{algorithm}[h!]
    \caption{The make\_global\_numbering routine, with our changes added}
    \label{updated-make-global-numbering}
    \begin{algorithmic}
        \Require local\_sec, global\_sec \Comment{local\_sec coming from create\_section} 
        \State val $\gets$ [ \ ] \Comment{holds the LGMap Indices} 
        \State pStart, pEnd $\gets$  chart for lsec
        \For{p in [pStart, pEnd]} 
        \State Get dofs, constrained\_dofs  at p from local\_sec
        \State Get loff at p from local\_sec \Comment{local offset, assuming dofs $>$ 0 } 
        \State Get goff at p from global\_sec \Comment{global offset}
        \If{constrained\_dofs $>$ 0}
        \For{c in range(constrained\_dofs)}
            \State val[loff + c] = -1 
        \EndFor
        \Else
        \For{c in range(dofs)} 
            \State val[loff + c] = $\vert goff \vert + c$ 
        \EndFor
        \EndIf 
        \EndFor \\ 
        \Return val
    \end{algorithmic}
\end{algorithm}

As seen in Algorithm \ref{updated-make-global-numbering}, the constrained degrees of freedom in the section, created in \texttt{create\_section}, are used here. As the correct constrained degrees of freedom have been mapped out by the time \texttt{make\_global\_numbering} is called, the code in \texttt{make\_global\_numbering} can access these constrained degrees of freedom, rather than requiring knowledge on where the boundary nodes are at the Python level such as when creating a regular \texttt{FunctionSpace} object. 

Printing the local to global map for the \texttt{RestrictedFunctionSpace} shows that the indices containing $-1$ are indeed indices 15 to 24, meaning that the local to global map and the offsets match up.  \par
\newpage
In conclusion, when boundary markers are passed through to a \texttt{RestrictedFunctionSpace}, this causes the renumbering of the degrees of freedom corresponding to each mesh point, moving all constrained degrees of freedom to the end of the core and owned block. This reordering is then carried through and seen in the local to global map for the \texttt{RestrictedFunctionSpace}. However, all of this means that in Firedrake the restricted space is just a \texttt{FunctionSpace} with a different numbering. To cut off these degrees of freedom from the matrix, the size of the matrix will have to change, which will be discussed in the next section as PyOP2 is responsible for this. 
\subsection{Changes to PyOP2}\label{section-pyop2-changes}
When Firedrake assembles a matrix, PyOP2 is called to build a \texttt{Sparsity} object, which sets the correct size of the eventual matrix returned to the user. This means that PyOP2 also has to have some information about the restricted degrees of freedom passed to the \texttt{Sparsity} object, so that the size of the matrix is reduced by the number of constrained degrees of freedom. This is the reasoning for moving all constrained degrees of freedom to the end of the list. If this did not occur, when the PyOP2 matrix size is reduced in the constrained case the wrong degrees of freedom could be cut off, and there would still be identity rows in the matrix.

\begin{figure}[h!]
    \centering
    \includegraphics[trim={0.5cm, 4.7cm, 0.5cm, 0.9cm}, clip, scale=0.32]{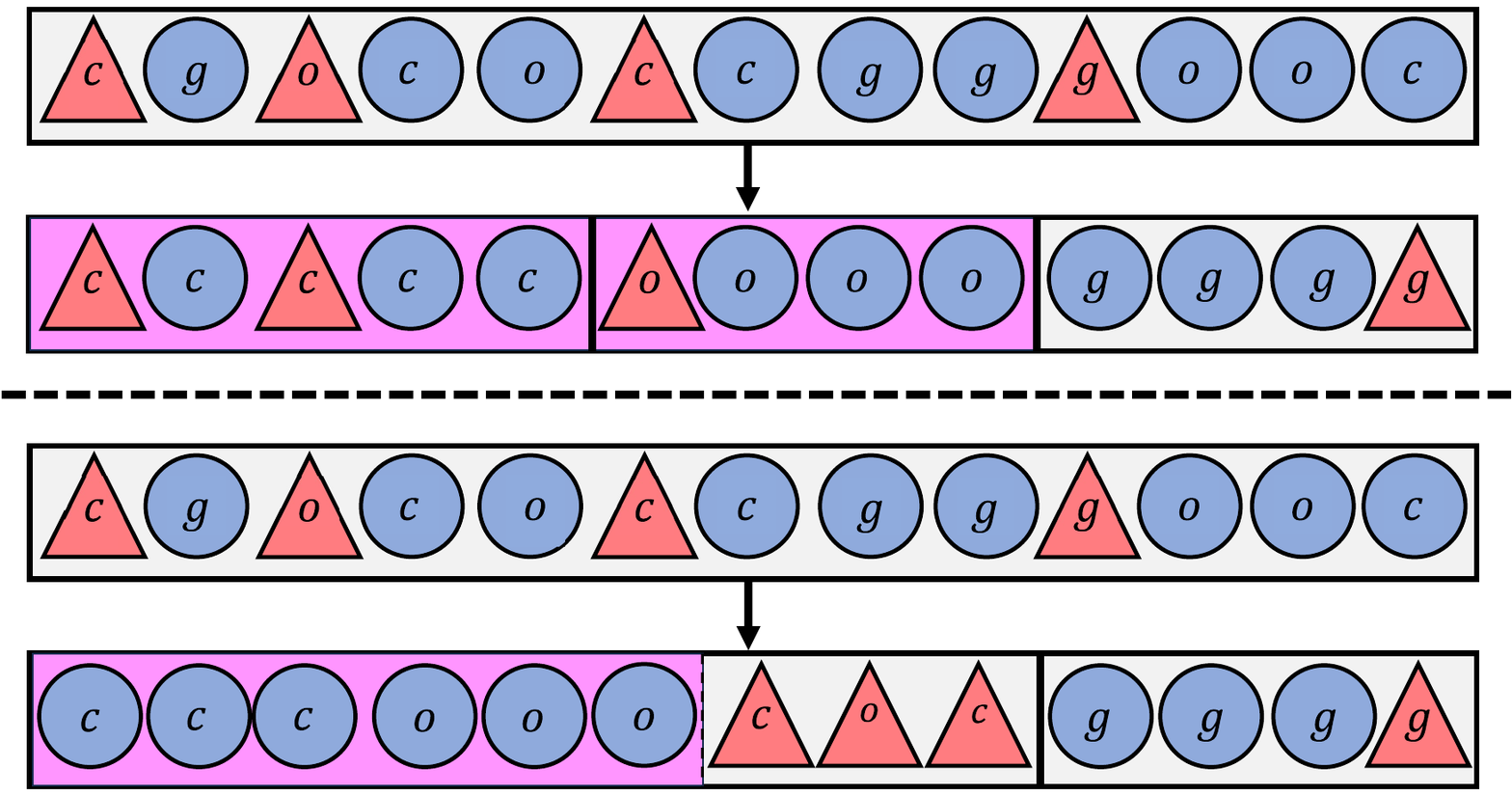}
    \caption{Figure \ref{plex-renumbering-diagram}, with purple highlights behind the mesh points that contribute to the assembled matrix size in the unrestricted (top) and restricted (bottom) cases. \vspace{-0.5cm}}
    \label{plex-renumbering-highlighted}
\end{figure}

Figure \ref{plex-renumbering-highlighted} shows which points contribute to the matrix size. In the restricted case, the \texttt{constrained\_size} parameter is 3, due to the three core/owned constrained mesh points. To achieve this removal of constrained degrees of freedom, a \texttt{constrained\_size} parameter is present in the constructor for the PyOP2 Set type, which records the number of constrained nodes. As this constrained size is the same as the number of constrained nodes found from the \texttt{Section} corresponding to the global numbering attribute in \texttt{FunctionSpaceData}, the number of constrained nodes is counted in \texttt{create\_section} and given as an output to \texttt{FunctionSpaceData} through the \texttt{get\_global\_numbering} function. 

When \texttt{FunctionSpaceData} is initialised, a function called \texttt{get\_node\_set} is called after the global numbering is created. This creates a PyOP2 \texttt{Set} object, with information about the nodes, called the \texttt{node\_set}. This \texttt{node\_set} is used in creating a \texttt{DataSet} object for the degrees of freedom, called the \texttt{dof\_dset}. A \texttt{DataSet} object inherits most of the properties of the \texttt{Set} that it is built on, such as the sizes of the set. The result of this is that it is sufficient to pass through the constrained nodes to PyOP2 solely when the \texttt{node\_set} is created. 

Algorithm \ref{get-node-set} shows the function \texttt{get\_node\_set}, which builds the node set as a PyOP2 \texttt{Set}. There is no new code, apart from the addition of constrained nodes, but it is useful to see where the new parameter is applied in the code.

\begin{algorithm}
    \caption{The \texttt{get\_node\_set} function of \texttt{FunctionSpaceData}. This shows how the global numbering as created in \texttt{create\_section} helps create the \texttt{node\_set} and eventual \texttt{dof\_dset}.}\label{get-node-set}
    \begin{algorithmic}
        \Require mesh, key 
        \State nodes\_per\_entity, real\_tensorproduct, boundary\_set $\gets$ key
        \State global\_numbering, constrained\_nodes $\gets$ get\_global\_numbering(mesh, key)  \Comment{effectively a wrapper for create\_section}
        \State Obtain correct node\_classes, halo \Comment{required for Set, not relevant here}
        \State node\_set $\gets$ PyOP2.Set(node\_classes, halo, mesh.comm, constrained\_size=constrained\_nodes)
        \State Check sizes \\ 
        \Return node\_set
    \end{algorithmic}
\end{algorithm}

This \texttt{constrained\_size} parameter as seen in the construction of a PyOP2 \texttt{Set} is only used when passed into the layout vector of the \texttt{dof\_dset}. The layout vector has a size, which is used as the size of the assembled matrix. The size of the layout vector is given as the following, defining the size of the node set to be the number of core and owned nodes:
\begin{equation*}
    (\text{size of node\_set} - \text{constrained\_size}) \times \text{dof per node}
\end{equation*}
This leaves existing \texttt{FunctionSpace} implementation unchanged, as the constrained size will always take the value of $0$ in this case. It also gives the correct calculation for the number of constrained degrees of freedom, except for special cases such as in elements where only some degrees of freedom on a node are constrained. 

\subsection{Changes to UFL}\label{section-ufl-changes}
A key issue was trying to distinguish between the \texttt{FunctionSpace} and \texttt{RestrictedFunctionSpace} classes, when creating certain objects such as the global kernel for use in assembly. As mentioned already, it was trivial to distinguish between the two within Firedrake, as we could just add the \texttt{boundary\_set} to cache keys used within the \texttt{FunctionSpaceData} object. This is seen, for example, in Algorithm \ref{get-node-set}. 

However, another issue was that some functions call the \texttt{ufl\_function\_space()} rather than Firedrake's own \texttt{FunctionSpace} / \texttt{RestrictedFunctionSpace} object. As we had only worked on Firedrake so far, UFL was unaware of the differences between restricted and unrestricted spaces due to the constructor for the UFL version of a \texttt{FunctionSpace} only taking in a domain and an element. 

When the code in Firedrake is assembling a form, it creates a PyOP2 \texttt{GlobalKernel} object within the \texttt{assemble} function. This is a cached object, so there is a check to verify if the function has been called before with the same arguments, and return the previous answer if so. When \texttt{assemble} was called on the same form, one made in the restricted space and one in the other, this cache was hit. In other words, the two inputs supposedly produce the same \texttt{GlobalKernel}. This should not happen, as the global numbering is different and so the  restricted space is distinct to the original one. To get around this, such as in the \texttt{FunctionSpaceData} cache, there was a need to add something unique to the cache key. This is to allow PyOP2 to separate the restricted and unrestricted case, removing cache errors. The addition to the key is the UFL form signature, as the procedure is assembling a form, so it makes sense to add a property of the object to be assembled rather than something further up the order of hierarchy such as the overall function space for one of the arguments within the form.
\begin{figure}[h!]
    \centering
    \includegraphics[trim={2.5cm, 11cm, 3.5cm, 0.5cm}, clip, scale=0.5]{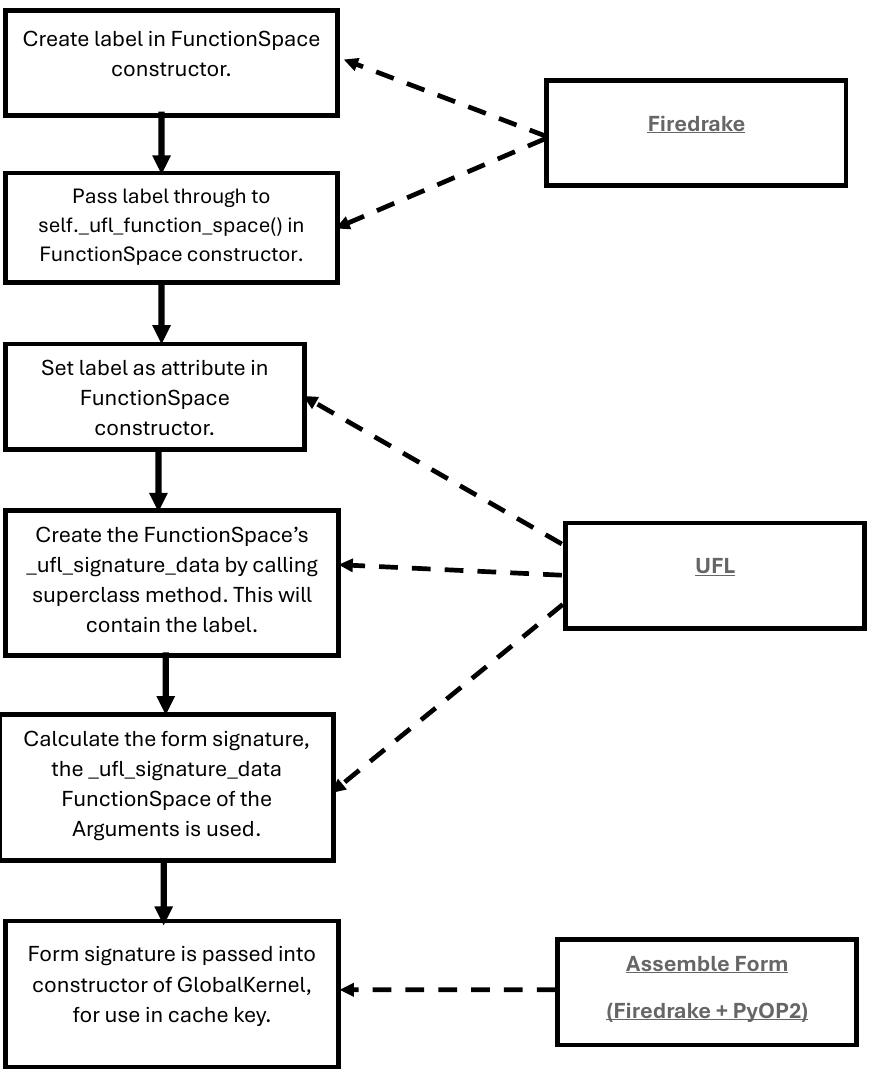}
    \caption{Places where the label added to a function space gets used. The left shows the specific use of the label, and the right shows where this use occurs. \vspace{-0.5cm}}
    \label{ufl-form-signature-traversal}
\end{figure}

To do this, there is now a ``label" attribute in the \texttt{FunctionSpace} and \texttt{RestrictedFunctionSpace} classes in both Firedrake and UFL. This contains the string representation of the boundary domains, meaning any \texttt{FunctionSpace} has an empty string but a \texttt{RestrictedFunctionSpace} does not. Figure \ref{ufl-form-signature-traversal} shows how this label gets passed through Firedrake to UFL, and then to eventually end up in the form signature. 


As the unrestricted function space has no boundary set, this should mean that initialising a \texttt{RestrictedFunctionSpace} with an empty boundary set should give the same results when assembling forms as well. 

\subsection{Changes in Parallel}\label{section-parallel-changes}
So far, most of the changes that have been seen in this section have been implicitly discussed in the context of serial computing. Of course, a lot of Firedrake users will want to use these new features in parallel. It turns out that the above algorithms have minor to no changes in the parallel case, as shown by the fact that none of them use the size of the MPI communicator which handles processing in parallel. For example, it is possible to extend Figure \ref{plex-renumbering-diagram} to show what happens in the parallel case with two process.

However, there is a change in how data is exchanged between processes, that is not covered in the serial case due to there being no data exchanges taking place. In the PyOP2 section, there was some discussion on how the mesh points get labelled through the use of Figure \ref{pyop2-labelling-cg4-unrestricted-2-processes} and what this means when data gets exchanged. The halo data exchange in Firedrake is implemented through a PETSc SF object, which led to an issue arising. 
\begin{figure}[h!]
    \centering
    \includegraphics[trim={3cm, 2.75cm, 2.5cm, 3.5cm}, clip, scale=0.4]{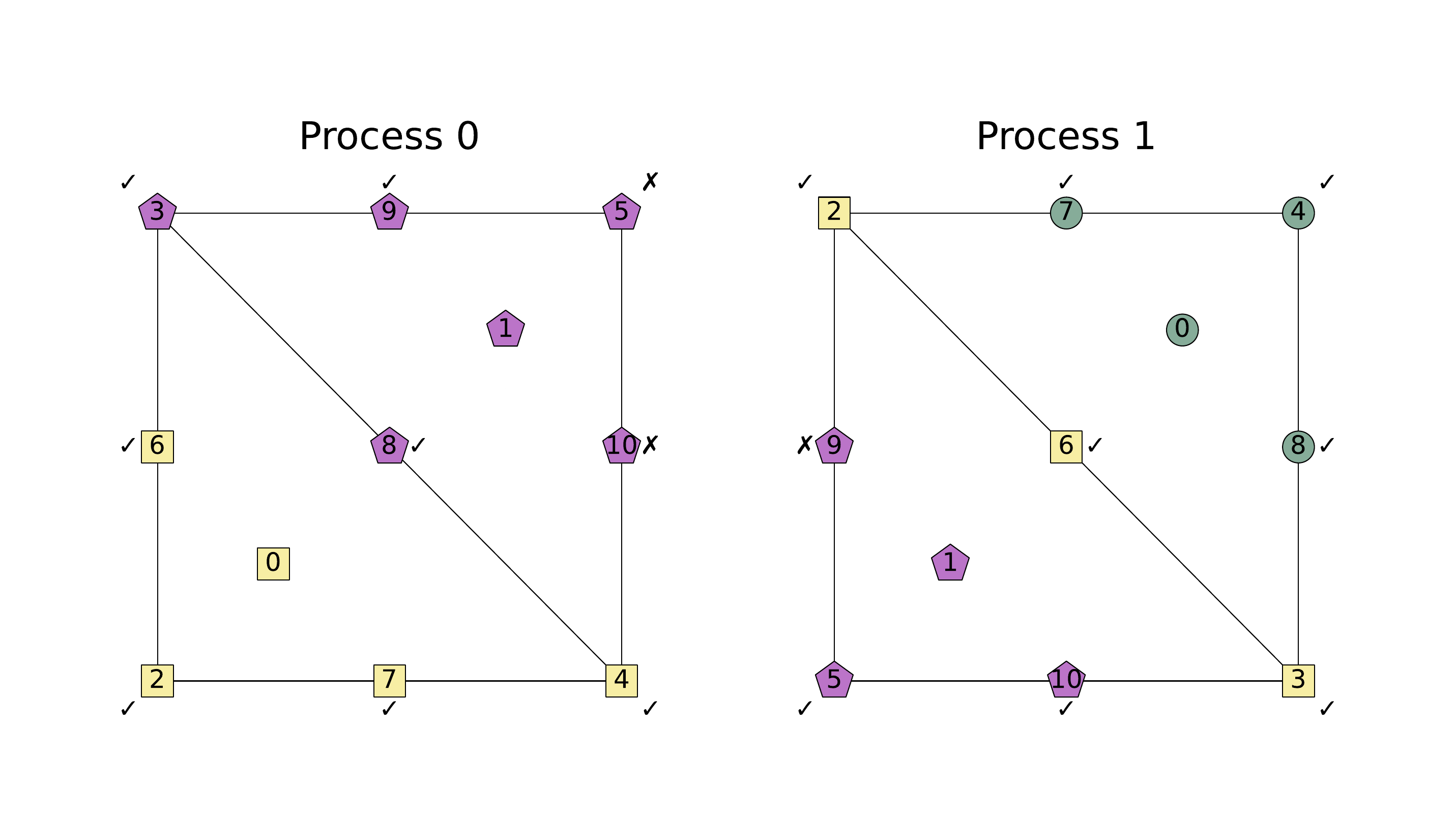}
    \caption{The labelling of a mesh when a \texttt{RestrictedFunctionSpace} of a Lagrange element of degree 2 with boundary conditions on the left and right boundaries of the square are defined. Using the same colour scheme as Figure \ref{pyop2-labelling-cg4-unrestricted-2-processes}, the green circles represent core points, yellow squares owned points and purple pentagons ghost points. Additionally, the ticks or crosses represent whether the coordinates that come out when interpolating $x$ or $y$ into a function on the \texttt{RestrictedFunctionSpace} are correct to the mesh, to illustrate the issues. \vspace{-0.5cm}}
    \label{numbering-issues-parallel}
\end{figure}

When marking degrees of freedom as constrained, the values on the constrained ghost points were not being updated in the data exchange, leading to some interesting results such as coordinates of distinct mesh points being set to $[0, 0]$ rather than their correct coordinate based on the layout based on the DM labels. In Figure \ref{numbering-issues-parallel} a reproduction of one of the initial diagrams that was drawn to try to isolate and fix the problem is presented. 

The constrained ghost points not being updated is somewhat challenging to see through this diagram, as some of the values are indeed correct. However, it is possible to see that indeed the only incorrect values come from the constrained ghost points in either process.

To debug, the values of the outputs of the \texttt{plex\_renumbering} and \texttt{local\_to\_global\_map} functions for the two processes can be displayed to see if the numbering of mesh points and offsets between these functions and the output of the \texttt{DM} as observed by Figure \ref{numbering-issues-parallel} were different. The output during the debugging stage appeared to be consistent, which was not the expected outcome of the debugging. The interested reader can look at Appendix A, (Section \ref{section-appendix-a}), for the details of what it meant for the output to be consistent, as there are multiple pieces of information to consider. 

Fortunately, Dr Koki Sagiyama provided an update to the function that creates the star forest used in halo data exchanges to remove this behaviour by manually creating the graph that the star forest object will read. This additionally uses the degrees of freedom at each point, rather than just the constrained degrees of freedom, so now the constrained points are included in the star forest too, and so any data exchange also updates these points. 

\subsection{Firedrake User-Level Changes}\label{section-user-layer}
In the previous sections, the description of how \texttt{RestrictedFunctionSpace} is implemented has been detailed. As mentioned, this section will focus on the features that optionally use \texttt{RestrictedFunctionSpace} in common functions, to make solving problems better for the user. 

To begin with, the \texttt{LinearEigenproblem} class is presented, which has a \texttt{restrict} parameter. As the name suggests, this parameter is a boolean to choose if the user wants to use restricted spaces when solving the problem. When \texttt{restrict = True}, all the \texttt{FunctionSpace} objects that are the spaces that the arguments are defined on in the forms will be switched to the equivalent restricted spaces using the supplied boundary conditions. The reproduction of the constructor for the LinearEigenproblem class, with some slight changes for clarity, is shown in Algorithm \ref{eigenproblem-setup}.

\begin{algorithm}[h!]
\caption{Algorithm to create a \texttt{LinearEigenproblem}, with \texttt{restrict=True}}
\label{eigenproblem-setup}
    \begin{algorithmic}
        \Require{restrict==True} \Comment{restrict==True is now the default}
        \Require{A, M=None, bcs} \Comment{Assume at least one boundary condition} 
        \State v, u $\gets$ A.arguments()
        \If{M}
        \State self.M = M
        \Else
        \State self.M = inner(u, v) * dx
        \EndIf
        \State self.output\_space = u.function\_space() 
        \State sub\_domains $\gets \bigcup_{i} bcs[i].sub\_domain$ 
        \State V\_res $\gets$ RestrictedFunctionSpace(self.output\_space, boundary\_set=sub\_domains) 
        \State u\_res, v\_res $\gets$ TrialFunction(V\_res), TestFunction(V\_res) 
        \State self.M $\gets$ replace(self.M, u $\rightarrow$ u\_res, v $\rightarrow$  v\_res) 
        \State self.A $\gets$ replace(A, u $\rightarrow$ u\_res, v $\rightarrow$  v\_res) 
        \State Replace boundary conditions to ones defined on the restricted space
        \State self.restricted\_space = V\_res
    \end{algorithmic}
\end{algorithm}

The replace function used in Algorithm \ref{eigenproblem-setup} is a function from the UFL package, The form taken in such as $A$ or $M$ in this case, has its arguments searched through a dictionary. If a function $u$ is in the keys of the dictionary, this function gets replaced by $dict[u]$ (another function). When applying the replace function in Algorithm \ref{eigenproblem-setup}, the replacements are from unrestricted to restricted spaces. 

What the code aims to achieve is the following change between the statements of the eigenproblem, using $A(u, v)$ to denote the matrix assembled from the bilinear form a(u, v) which gets passed into the \texttt{LinearEigenproblem}.
\begin{align*}
    \text{Original Problem: Find } u \in V \text{ such that } A(u, v)u = \lambda M(u, v)u \ \forall v \in V \\
    \text{New Problem: Find } u \in V_0 \text{ such that } A(u, v)u = \lambda M(u, v)u \ \forall v \in V_0
\end{align*}
In Algorithm \ref{eigenproblem-setup}, there is no use of a shift of $\theta$, which was described in the motivation section. In Firedrake, there is still the option to choose to take a value of $\theta$ when defining a \texttt{LinearEigenproblem}, which defaults to 0. If $restrict==True$, this $\theta$ value is ignored, due to the fact there are no longer any boundary eigenvalues, so this parameter is omitted from the algorithm for clarity. 

The data associated with the eigenvectors, which are presented as an output to the user, are functions in the original space - not the restricted one. This is to prevent confusing users who are expecting to solve a problem on one space, and instead are receiving an interpolated version of the results they expect, due to the change to the restricted space. This is done when the user calls \texttt{eigenfunction} on a \texttt{LinearEigensolver} class that has already been used to solve a problem, by interpolating the results from the restricted space back to the original space. This interpolation should be relatively simple, as the previous sections have shown the difference between the restricted and unrestricted space is the renumbering of the same degrees of freedom. 

For the changes in the top level, it is the use of the \texttt{replace} function from UFL that allows arguments of a form to be swapped so that they are defined on a restricted space rather than an unrestricted one. Due to these changes to the eigensolver, the tutorial under the ``Oceanic Basin Modes: Quasi-Geostrophic approach" heading in Firedrake's user manual \cite[pp. 122-127]{FiredrakeUserManual} has been updated. This uses the eigensolver to solve a problem, and an additional note is now present to show that this eigensolver uses $restrict==True$ in the definition of the eigenproblem.

Additionally, the same ideas as detailed in the eigensolver are also present in the \texttt{solve} function, through a \texttt{restrict} parameter. This process is almost the same as the process in the eigensolver, with the switching of spaces in the forms required to the restricted version but still returning the output as a function of the unrestricted space. Different forms have changes made to them when calling \texttt{solve}, such as the Jacobian of the variational problem as well as the residual, but as the ideas are so similar the algorithms used are not reproduced. 

As a final user-level change, there is a warning raised in two cases where the user is using a \texttt{RestrictedFunctionSpace} incorrectly:
\begin{itemize}
    \item The user defines the boundary set such that they intend to restrict every single degree of freedom. If this occurs and is intentional, the user wants to set the value of every node with a known boundary condition, so there is no point in solving a problem in the first place. Therefore, it is more likely that an error is made, so the user is warned when they do this. 
    \item The user creates a \texttt{RestrictedFunctionSpace}, but then goes on to define a boundary condition on the space on a subdomain that does not exist in the boundary set. This is not allowed, as the \texttt{RestrictedFunctionSpace} is already created and so will have already done the work to mark the nodes based on the boundary set as constrained by the time the boundary condition is defined. This would give the wrong results, as the spaces thinks the restriction is on one area but the boundary condition is trying to restrict another, so the user is once again warned when they do this. Otherwise, a result of an assembly might end up erroneously containing boundary rows, using the small cases below as an example.
    \begin{equation*}
    `   \begin{pmatrix}
            1 & 0 & 0 & 0 \\ 
            0.5 & -0.5 & 1 & 2 \\ 
            0 & 0 & 1 & 0 \\ 
            0.3 & 0.7 & -0.8 & 2
        \end{pmatrix}\rightarrow\begin{cases}\begin{pmatrix}
            1 & 0 \\ 
            0.5 & -0.5
        \end{pmatrix} \text{mismatch: boundary set = (1, 3), condition on (3, 4)}\\ 
        \begin{pmatrix}
            -0.5 & 2 \\ 
            0.7 & 2
        \end{pmatrix} \text{match: boundary set = (1, 3), condition on (1, 3)}
        \end{cases}
    \end{equation*}
\end{itemize}
With these warnings in place, errors using the \texttt{RestrictedFunctionSpace} on the user side should be minimised. There is potential to go even further to add in error warnings or automatic changes when trying to solve a restricted problem with the restrict flag, but these should not be issues that would cause unexpected solutions and too many errors and exceptions may make the code hard to use.

\section{Testing}\label{section-testing}
The code of Firedrake has a directory for tests of major components, with the aim of ensuring that each piece of code works as expected. These tests run whenever a new update is pushed to Firedrake, to also ensure new additions do not break the functionality of the rest of the current code. When developing the \texttt{RestrictedFunctionSpace} class, two types of test were added to the code base: one type that tests the solving capabilities when using the space directly and the other to test the top-level flags. 

A first test of the \texttt{RestrictedFunctionSpace} class is to take a unit square mesh, with a boundary condition on one side of the square, and determine if the matrix from an assembled form defined in a restricted space has the correct shape and entries.
The finite element that is used in the test is the Lagrange element of degree 1, which has one degree of freedom at each corner of the square mesh. As one side of the mesh has a boundary condition applied, 2 of the 4 degrees of freedom disappear when using the appropriate restricted space. When forming the matrix of Equation \ref{matrix-vector-system} using \texttt{assemble} in Firedrake, when using the normal \texttt{FunctionSpace} class a $4\times4$ matrix with identity rows is returned. The form defined in the \texttt{RestrictedFunctionSpace} assembles a $2\times2$ matrix without the identity rows and columns, with the same non-identity entries as the matrix using the \texttt{FunctionSpace}. Evidently, this is the correct behaviour, as the identity rows and columns belong to degrees of freedom on the boundary, and we expect 2 of the 4 degrees of freedom to be constrained. This test is described in Algorithm \ref{test-assembly}.

\begin{algorithm}
    \caption{Test a small assembly of a form in a RestrictedFunctionSpace on UnitSquareMesh(1, 1)}
    \label{test-assembly}
    \begin{algorithmic}
        \State mesh $\gets$ UnitSquareMesh(1, 1)
        \State x, y $\gets$ SpatialCoordinates(mesh)
        \State V $\gets$ FunctionSpace(mesh, ``CG", 1) 
        \State bc $\gets$ \texttt{DirichletBC}(V, 0, 1)
        \State V\_res $\gets$ RestrictedFunctionSpace(V, [bc])
        \State original\_form $\gets u \times v \ dx$ \Comment{u = TrialFunction, v = TestFunction}
        \State restricted\_form $\gets u_{res} \times v_{res} \ dx$ 
        \State restricted\_fs\_matrix $\gets$ assemble(restricted\_form)
        \State fs\_matrix $\gets$ assemble(original\_form, bcs=[bc])
        \State fs\_matrix $\gets$ Removed identity rows from fs\_matrix \\ 
        \Assert fs\_matrix == restricted\_fs\_matrix
    \end{algorithmic}
\end{algorithm}

Using the same structure and assertions as Algorithm \ref{test-assembly}, different orders of the Lagrange element on a unit square mesh are tested, as well as tests occurring for the Lagrange element on meshes with different number of elements in the $x$ and $y$ directions. The Lagrange element is the element that is used for the majority of testing, as most of the code added is in places that are written independently of element type, with the notable exception of the Hermite element. \par

Tests are also carried out to ensure that when \texttt{solve} is called, the \texttt{RestrictedFunctionSpace} returns the correct solution. This is based on confidence that the left-hand side of the matrix-vector system is correct and so the solution should be correct. Some known solutions are compared to the numerical solution, while others are tests comparing the results coming from the unrestricted and corresponding restricted space. 

Additionally, a test is carried out for \texttt{MixedFunctionSpace}, with the test problem using the mixed formulation of the Poisson problem $-\Delta u  = f$, as described in the manual \cite[pp. 95-97]{FiredrakeUserManual}. The problem is shown below, using the notation of $\sigma = \nabla u$:
\begin{align}
    &\sigma - \nabla u = 0 \text{ on $\Omega$} \label{eq-1-mixed} \\ 
    &\nabla \cdot \sigma = - f \text{ on $\Omega$} \label{eq-2-mixed}\\ 
    &u = u_0 \text{ on $\Gamma_D$} \label{eq-3-mixed}\\ 
    &\sigma \cdot n = g \text{ on $\Gamma_N$} \label{eq-4-mixed}\\ 
    &f = 1, g = 0, u_0 = 0, \Gamma_{N} = \{(x, 1) \in \partial\Omega\}, \Gamma_{D} = \partial\Omega - \Gamma_{N} \nonumber
\end{align}

This problem can be transformed into the variational problem by creating a function space $W$ = $V_\sigma \times V_u$, and multiplying Equation \ref{eq-1-mixed} by $\tau \in V_\sigma$ and Equation \ref{eq-2-mixed} by $v \in V_u$. Integration by parts then gives us \cite[pp. 95-97]{FiredrakeUserManual}:
\begin{equation}
    \label{mixed-poisson-var-form}
    a((\sigma, u), (\tau, v)) = \int_{\Omega} \sigma \cdot \tau + \nabla \cdot \tau u + \sigma \cdot \sigma v dx,  \ \ G((\tau, v)) = - \int_{\Omega} fv dx + \int_{\Gamma} u \tau \cdot n ds 
\end{equation}
In this case, we see that $g$ must be a Dirichlet boundary condition as it does not appear in Equation \ref{mixed-poisson-var-form}, whereas $u_0$ is a Neumann boundary condition, through the $\int_{\Gamma} u \tau \cdot n ds $ term in $G$. For the \texttt{MixedFunctionSpace}, the test solves for both $\sigma$ and $u$ through the process of multiplying two function spaces together. This is done using both restricted and unrestricted spaces, and the values of the subfunctions found are compared, to assert that the norm of the error norm between the two solutions is sufficiently close. 

Another test for the \texttt{RestrictedFunctionSpace} is to build and solve a problem in different coordinate systems. This test was motivated by some initial testing during the construction of the class showing that we could correctly solve problems with the Lagrange element of degree greater than 1, but not of degree 1 itself. Once that problem was resolved, through defining separate form signatures for restricted and unrestricted spaces, this test also passes as expected. 

An important top-level test is the test of the updated eigensolver. This test occurs in two ways, one is to ensure the restrict flag had the same behaviour as using a \texttt{RestrictedFunctionSpace} in the forms that initially define the \texttt{LinearEigenproblem}, and the other is to ensure the $restrict==True$ and $restrict==False$ results were similar to each other. The test was constructed from a similar problem to the oceanic basin problem in the tutorial \cite[pp. 122-127]{FiredrakeUserManual}.

An additional test, which was suggested and created by Dr Koki Sagiyama, is to use the same style of test for the updated solve function and run the test in parallel. This is to ensure that \texttt{RestrictedFunctionSpace} produces correct results both in serial and parallel.

\section{Comparison to FunctionSpace}\label{section-comparison}
In this section, the two function space classes in Firedrake, \texttt{FunctionSpace} and \texttt{RestrictedFunctionSpace} are compared. The objective of this chapter is to show that the \texttt{RestrictedFunctionSpace} performs similarly to the \texttt{FunctionSpace} with respect to accuracy of solutions, while retaining the benefits that were theoretically given through the creation sections. 
To illustrate the differences between the behaviour of \texttt{RestrictedFunctionSpace} and \texttt{FunctionSpace} when solving a variational problem, Firedrake is used to solve the problem used throughout the paper, Poisson's equation (Equation \ref{poisson-equation}). However, there is only a boundary condition on the $x=0$ boundary specified on the square, rather than on both $x=0$ and $x=1$, so that the choice in function space does not lead to constraining all the available nodes. 

For this demonstration, the matrix for the bilinear form is assembled using the Lagrange element of degree 1. This is defined analogously to the element of degree 4, where each triangle has one degree of freedom at each corner and no more. This means that the assembled matrix on the mesh will be of size $4 \times 4$ on the \texttt{FunctionSpace} and of size $2 \times 2$ on the \texttt{RestrictedFunctionSpace}, as the two degrees of freedom on $x=0$ are restricted. The code that assembles this matrix is very similar to Listing \ref{poisson-code} mentioned in the introductory sections of the report, with the change of $4$ to $1$ in the declaration of the \texttt{FunctionSpace}, and the removal of the right boundary condition. 
To create a \texttt{RestrictedFunctionSpace} from the space $V$, restricting on $x=0$ the following line of code is used. 
\begin{minted}{python}
    V_res = RestrictedFunctionSpace(V, boundary_set={1})
\end{minted}
When using a \texttt{RestrictedFunctionSpace} within the code, the assembly steps remain the same, but all test and trial functions and boundary conditions will be created using this space instead of the \texttt{FunctionSpace}. 
\par 
The matrices returned from the assembly of the bilinear form associated with the Poisson problem is shown in the equation below. The left matrix corresponds to the unrestricted space, and the right matrix is created using the restricted space. Both matrices have the entries rounded to 4 decimal places for legibility. 
\begin{equation*}
    \begin{pmatrix}
        1 & 0 & 0 & 0 \\ 
        0 & 0.0833 & 0.0417 & 0 \\ 
        0 & 0.0417 & 0.1667 & 0 \\ 
        0 & 0 & 0 & 1 
    \end{pmatrix}, \ \ \  
    \begin{pmatrix}
        0.0833 & 0.0417 \\ 
        0.0417 & 0.1667 
    \end{pmatrix}
\end{equation*}

We can immediately see the difference in the size of the matrix assembled between the two spaces, and that the matrix produced using the \texttt{RestrictedFunctionSpace} has no identity rows or columns, which is as expected. Now that this example has shown that there is consistency between the matrices assembled in the two spaces, a variational problem is solved to show the solutions produced in the spaces are also the same. The consistency between solutions can be shown in two different ways: comparison when using the ``restrict" flag as True/False in the call to \texttt{solve}, and one way using a \texttt{RestrictedFunctionSpace} in the definition of the forms and functions going into the \texttt{solve} function compared to using a \texttt{FunctionSpace}. 
\par

\begin{figure}[h!]
    \centering
    \includegraphics[trim={4cm, 1cm, 1cm, 1cm}, clip=True, scale=0.3]{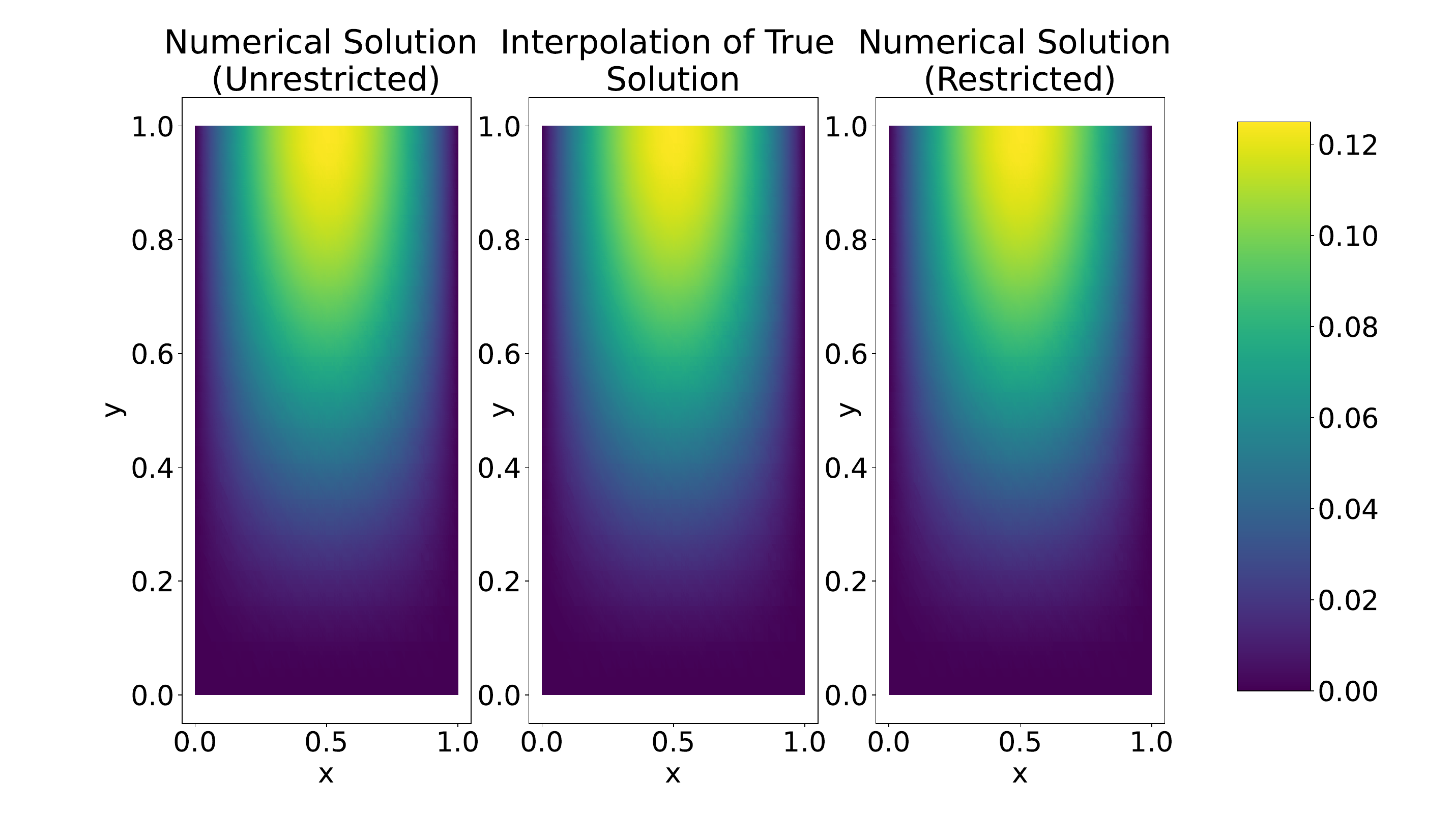}
    \caption{Solution to $-\Delta u = -2(y^3 - y^2) + (x-x^2)(6y - 3)$ in Firedrake. Left is the numerical solution on an unrestricted space, Centre is the true solution given by Equation \ref{u-solution-compare-fs-rfs}, and Right is the numerical solution on a restricted space. Each solution is created using a mesh with 16 cells.}
    \label{solution-plot-3-ways}
\end{figure}

To do this, we will solve a problem that we have the analytic solution for, to cross-reference both the restricted and unrestricted solutions with the exact solution. This problem is created using the method of manufactured solutions \cite[pp. 98-99]{finite-element-course}, which is a way to produce partial differential equations given a known solution.
If we take Poisson's equation in the unit square, $-\Delta u = f$, with homogeneous Dirichlet boundary conditions on the left and right-hand side, and homogeneous Neumann boundary conditions on the top and bottom, we can choose a certain known function for $u$ satisfying these boundary conditions. One such function $u(x, y)$ that does this is the following:
\begin{equation}
\label{u-solution-compare-fs-rfs}
    u(x, y) = -(y^3 - \frac{3}{2} y^2)(x)(1-x) 
\end{equation}
Now we have a solution, we fit $f$ to match the solution, by using $-\Delta u = f$. We therefore get:
\begin{equation*}
    f = -2(y^3 - y^2) + (x-x^2)(6y - 3)
\end{equation*}
The solution in the restricted space can be created through the same method as the code used in Listing \ref{poisson-code}, with a slight change to the code in that additionally the code will include a line to define and interpolate $f$ into the chosen space.  It is possible to qualitatively determine if the solutions are equal by referring to a heatmap of the solution, such as the one shown in Figure \ref{solution-plot-3-ways}. In this figure, it is clear that there is no difference between the exact solution, the unrestricted solution or the restricted solution, suggesting the restricted function space is working correctly when solving a problem.

The mesh is now divided into 16 cells in the $x$ and $y$ direction, rather than just one, and Lagrange elements of degree 2 are used so that sufficiently detailed plots of the solution are produced. For a numerical check, the value of the error norm between the output and the solution found in Equation \ref{u-solution-compare-fs-rfs} is obtained. This error norm is $4.20 \times 10^{-7}$, which is quite small but could also be improved with a more optimal choice in mesh or function space. 
The solution in both cases should be identical, up to a rearranging of the nodes. To confirm that this is the case, the solution in the restricted space can be interpolated into the unrestricted space, which occurs when the \texttt{errornorm} function calculates the error norm between the two numerical solutions. This value is $3.13\times 10^{-16}$, which is small enough to confidently say that the two solutions are the same.

\begin{figure}[h!]
    \centering
    \includegraphics[trim={1.5cm, 0.7cm, 0.5cm, 0.65cm}, clip, scale=0.32]{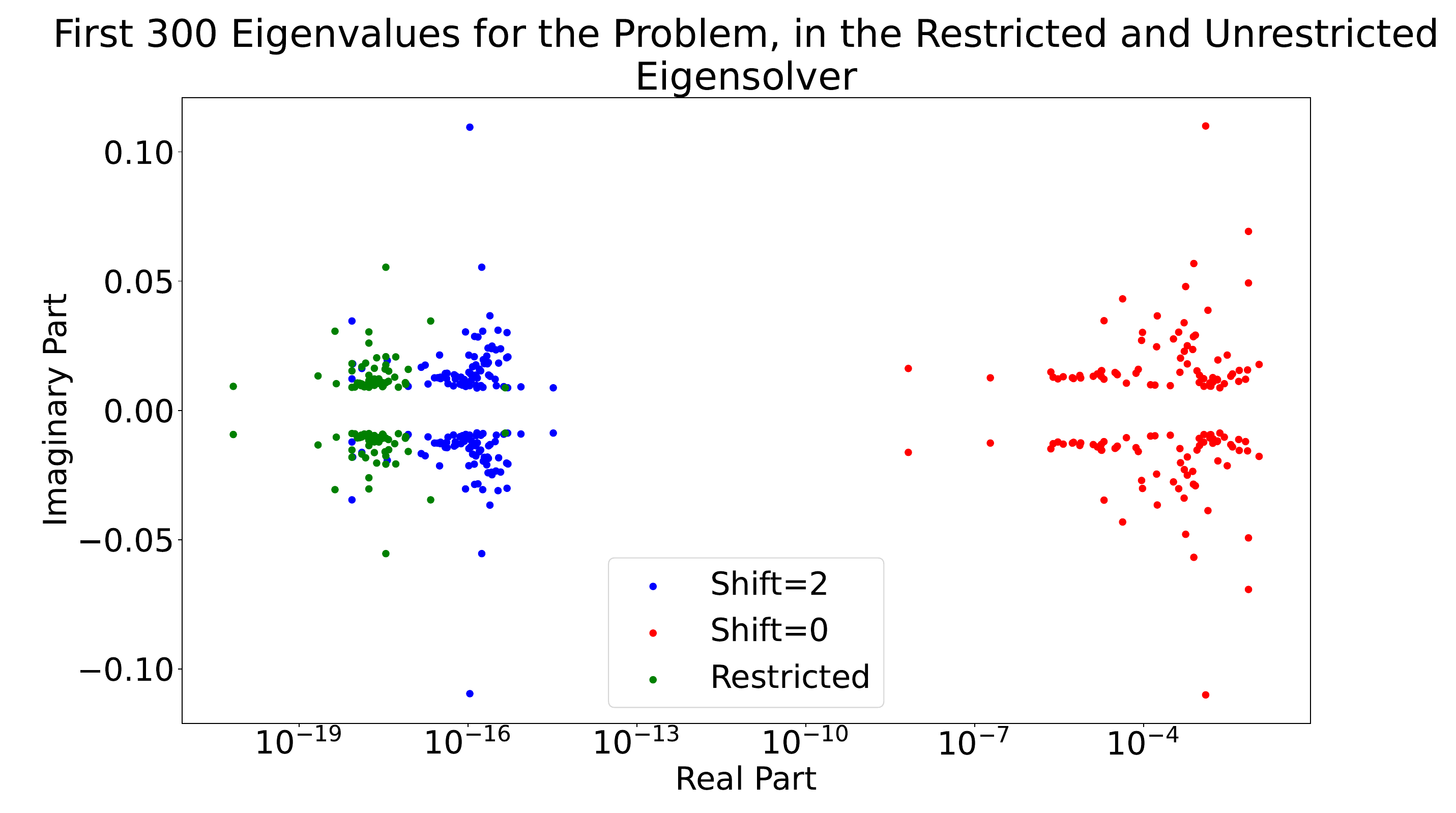}
    \caption{Solution to the problem given in Equation \ref{oceanic-basins-eigenproblem}, but with the \texttt{restrict=True} flag enabled in the LinearEigenproblem (green). The eigenvalues found when restrict=False for two different possible shifts are also given for comparison. \vspace{-0.3cm}}
    \label{bcshift-2-0-restrict!}
\end{figure}

In Section \ref{section-motivation}, there was discussion about how Firedrake currently shifts the eigenvalues of the boundary conditions so that a clash in eigenvalues does not occur and the issues that come from this. The oceanic basin model was described in that section, and it is shown in Figure \ref{bcshift-2-0-problem!} that a suboptimal choice in eigenvalue shift, such as $0$ for this problem, can affect the eigenvalues returned. In this section, the same problem is solved in the restricted space, by setting $restrict=True$ in the eigenproblem object. The first $300$ eigenvalues are shown in Figure \ref{bcshift-2-0-restrict!}, alongside the eigenvalues as found in Figure \ref{bcshift-2-0-problem!}.  \par 

These eigenvalues are close to, but not the exact same as, the eigenvalues found with the more optimal shift parameter of 2. As seen in the image, the eigenvalues that are returned from the eigensolver when \texttt{restrict=True} are very close in magnitude to the ones that are returned when the shift parameter is 2 and \texttt{restrict=False}. The explanation for the discrepancy could be as simple as the fact that the real parts all have a magnitude of around $10^{-17}$, which is close to machine precision, and the eigenvalues are in fact the same up to some small numerical error in the real part.

However, as these eigenvalues displayed in Figure \ref{bcshift-2-0-restrict!} are not exactly the same, to convince the reader of the correctness of the restricted eigenproblem for finding eigenvalues a problem with a known solution is also solved. The problem that is solved is the same as the problem used in the test that already exists in Firedrake for the eigensolver \cite{eigensolver-test}. This test uses the 1D Poisson eigenproblem on the interval $[0, \pi]$ given below \cite[pp. 5]{eigenproblem-book}.
\begin{equation}
    \label{poisson-eigenproblem}
    -u''(x) = \lambda u(x), \ \ u(0) = u(\pi) = 0 
\end{equation}

The Poisson eigenproblem in matrix form, which is $Ax = \lambda Mx$ is formed using the ideas of the variational form from the introduction to finite elements section. The only change between Equation \ref{poisson-equation} and Equation \ref{poisson-eigenproblem}, apart from the change from many dimensions to one, is the multiplication of the right-hand side by $u$ as well. This means that the form of $A$ and $M$ will be of the form as given below \cite[pp. 5]{eigenproblem-book}:
\begin{equation*}
    A_{ij} = \int^{\pi}_{0} \phi_j'(x)\phi_i'(x) dx,\ M_{ij} = \int^{\pi}_{0} \phi_j(x)\phi_i (x)
\end{equation*}
This can easily be compared to the matrix-vector system of Equation \ref{matrix-vector-system}, noting we are now in a matrix-matrix system. This problem has the known eigenvalues of $\lambda = n^2, n \in \mathbb{N}$ \cite[pp. 5]{eigenproblem-book}. Further results on the structure of the eigenfunctions are known, but they are not used here. 

\begin{figure}[h!]
    \centering
    \includegraphics[trim={4cm, 1cm, 2cm, 2cm}, clip, scale=0.3]{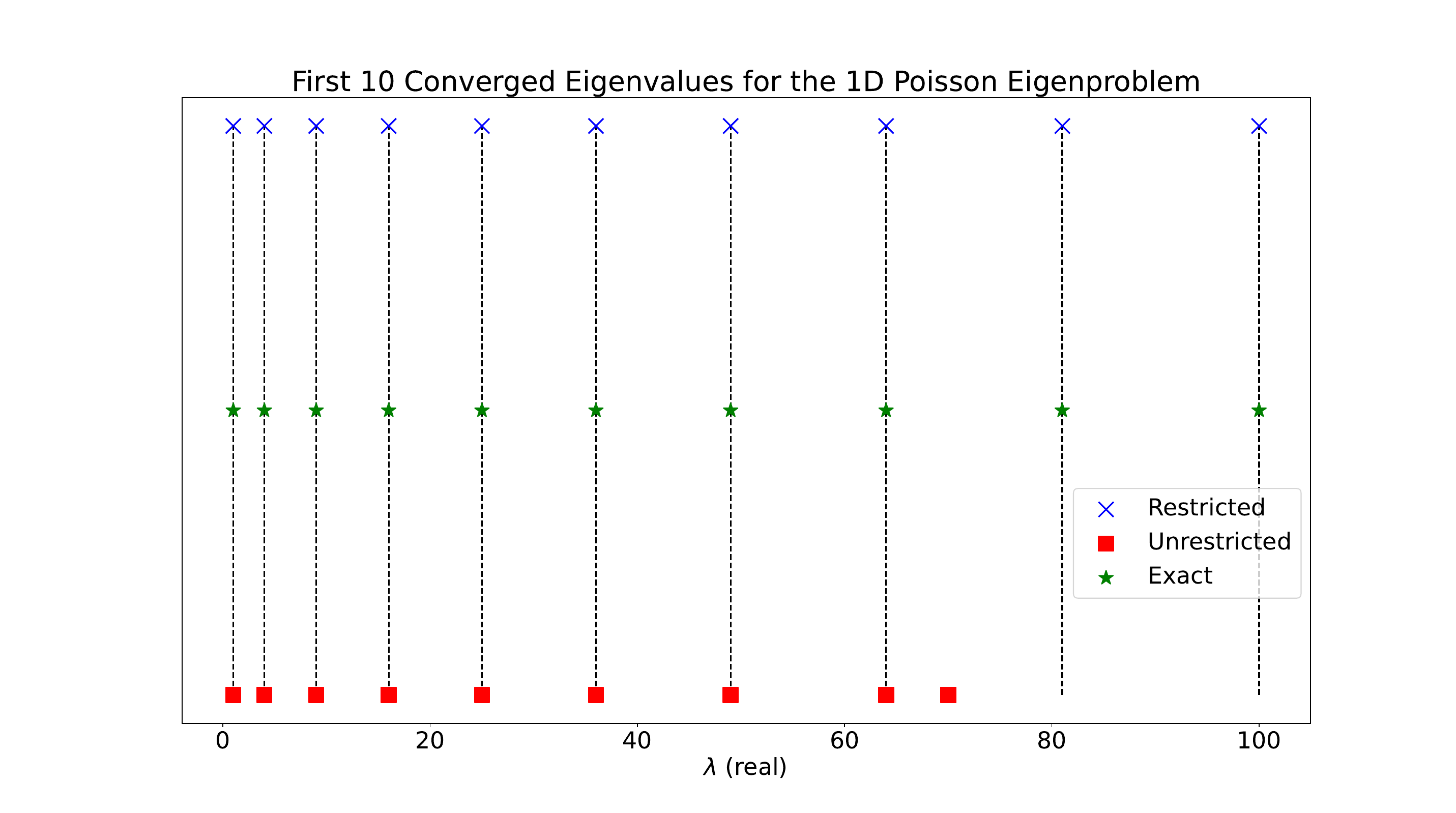}
    \caption{Eigenvalues for the 1D Poisson Problem. Top: eigenvalues found from the eigensolver, using the restricted function space. Middle: true eigenvalues, of the form $\lambda = n^2$. Bottom: eigenvalues found from the eigensolver, using an unrestricted function space and shifting boundary eigenvalues by $70$. \vspace{-0.3cm}}
    \label{eigenvalues-1d-poisson}
\end{figure}

To obtain the plot in Figure \ref{eigenvalues-1d-poisson}, a $4^{th}$ order Lagrange element is used on a mesh with 10 cells dividing the interval, when solving the eigenproblem. From Equation \ref{poisson-eigenproblem}, using the Lagrange element means that there are two boundary degrees of freedom on the interval, corresponding to the degree of freedom at $0$ and the degree of freedom at $\pi$.

When setting up the eigensolver, the code specifies that 10 eigenvalues should be found, although more than 10 eigenvalues are returned for certain shifts. Additionally, when using an unrestricted function space, the shift of the boundary condition eigenvalues was set to be $70$ to produce the figure. This shift shows clearly which eigenvalues are the boundary ones, as the solution to the problem is known. If the shift was set to be $n^2$, it would be impossible to distinguish between the boundary and non-boundary eigenvalues at that shift. A final conclusion of Figure \ref{eigenvalues-1d-poisson} is that the first 10 eigenvalues returned from the restricted eigensolver are all correct, which gives confidence to the validity of the results from the Oceanic basin eigenproblem.

In conclusion, the \texttt{RestrictedFunctionSpace} and the \texttt{FunctionSpace} both return the same answers when used in the code to solve problems, giving confidence to the validity of the implementation of the \texttt{RestrictedFunctionSpace}. However, it is evident that the \texttt{RestrictedFunctionSpace} performs much better when solving an eigenproblem, especially when an optimal choice in shift for the boundary condition eigenvalues is not known, as more eigenvalues and importantly more accurate eigenvalues are found. To further differentiate between the two spaces when solving a variational problem, it could be recommended to perform a timing test on a standard problem, to confirm whether the \texttt{RestrictedFunctionSpace} reduces or increases computational time.

\section{Conclusions and Further Work}\label{section-conclusion}
In conclusion, the implementation of the \texttt{RestrictedFunctionSpace} class has been successful in achieving the goals that were set out within this report. There is now a well-defined method in Firedrake to split the function space to focus on just the subset of functions which do not span the area on which Dirichlet boundary conditions are defined. In the following paragraphs, the evidence that the goals have been achieved are analysed, as well as any potential further work to improve \texttt{RestrictedFunctionSpace}. \par

A clear goal of any addition to a large library is to ensure that the user knows how to use this new feature. It is reasonable to say that the class is accessible to all users, through the built-in high-level features of the optional keyword flags in commonly used Firedrake functions. Additionally, for the more advanced user, the option to build a \texttt{RestrictedFunctionSpace} directly is there, allowing for a more transparent way to show the class is being used in code. Both approaches will have different advantages for different types of user, so should be effective no matter the use case. For additional information for the user, the class is documented with descriptions of the boundary set, and equivalently the functions using the restrict flag have this mentioned in the documentation. To further promote the class, the tutorial introducing the eigensolver capabilities that Firedrake has \cite[pp. 122-127]{FiredrakeUserManual} has been updated to include a brief explanation about the \texttt{restrict} flag. This, plus this report and any documentation on the website for Firedrake should be sufficient material for the users to understand and work with the changes.

There is confidence that the \texttt{RestrictedFunctionSpace} works in most situations that a user may use it in due to the amount of testing done on the class, coupled with the general lack of element-specific code in Firedrake. Furthermore, in the previous section of this report, there are some demonstrations of solving a variational problem using a \texttt{RestrictedFunctionSpace} and how the results compare to those produced when using a \texttt{FunctionSpace}. The goal of testing is so that the user can trust the functions they are using. Of course, some unforeseen and specific bugs may come up, and these should be explored and removed if possible. At the time of writing the report, there were no large bugs that have come to the attention of the author, but this will inevitably change as more uncommon and unexpected uses of the \texttt{RestrictedFunctionSpace} object occur. A non-exhaustive outline of some of the potential issues are given below, to flag them as future areas to focus on.

For an example of an issue that might appear, some initial use of the new flags at the top level to solve problems suggests that sometimes the performance of some of the built-in restrict flags to swap a \texttt{FunctionSpace} for a \texttt{RestrictedFunctionSpace} when solving a variational problem or eigenproblem might be slower compared to when the flag is not used. This was not tested in any official capacity during the project, so testing should occur before concluding this is definitely an issue. A likely cause of the issue may be due to some inefficiencies in the code, with a lot of replacements and building of new spaces and functions. These issues could be looked into, to be reproduced and dealt with formally so that the benefit of the restrict flag is seen in all aspects of the use of Firedrake rather than just the mathematical usefulness. \par

Additionally, in the code it was necessary to set \texttt{restrict=False} as the default behaviour \texttt{solve} function due to multiple errors caused when setting \texttt{restrict=True} as the default. These errors occur because a lot of element types have not yet been considered when creating a \texttt{RestrictedFunctionSpace}, and due to other complications due to \texttt{solve} being a very general function. For example, it is currently not possible to restrict a \texttt{FunctionSpace} that was created using Hermite elements on an interval. The issue is that every other degree of freedom on the end node is a boundary degree of freedom in the Hermite interval element, rather than every one for other elements, and this changes the numbering yet again. This lack of support for Hermite elements will be disappointing for some users, who choose to work that element for a specific reason and so cannot benefit from the additions that restricting the space could offer. A simple extension would be to look at how to calculate the restricted degrees of freedom for Hermite elements.  
Another function space type that is not currently explicitly supported when creating a \texttt{RestrictedFunctionSpace} is vector-valued function spaces. It is possible that the \texttt{RestrictedFunctionSpace} class works in the current state for vector spaces, but the tests in Firedrake do not test for this capability, and as such some bugs could have gone unnoticed. In defining vector-valued function spaces, there are two options. One option is to use \texttt{VectorFunctionSpace} object, which is a vector-valued space where each component is a \texttt{FunctionSpace} \cite[pp. 3]{FiredrakeUserManual}. The other method is to use a \texttt{VectorElement} and put this into \texttt{FunctionSpace} itself. These produce the same result, and so what works for one method should work with another.  

In a converse problem, there are a lot of elements that a user cannot define a \texttt{DirichletBC} objects on. Currently, the errors associated with this are called at the point of creating a boundary condition rather than the \texttt{RestrictedFunctionSpace}. It would make sense to add in a system checking the type of element at the creation of the \texttt{RestrictedFunctionSpace} to avoid this from happening, as it is redundant having a \texttt{RestrictedFunctionSpace} that does not create any boundary conditions.

These potential updates are important to consider, but there is the argument that trying to anticipate what users want is a losing position to be in. It is possible these updates that are suggested, such as the update to the Hermite element, are not relevant or even desired and the continuation of \texttt{RestrictedFunctionSpace} will be in a direction that is completely unforeseen as of now. A lot of future work will depend on what is currently going on right now in the development of Firedrake, and what features are being added to it, which is an evolving process. 

All of these possible future updates and the conclusions based off the implementation of the project are useful, and are what will be utilised the most, but there is one final question left to answer, to ensure the goal of the project was met. Did this change encapsulate the mathematics behind the finite element method better than what was there previously? Certainly, as heavily discussed throughout the report, \texttt{RestrictedFunctionSpace} is an attempt to remove some of the computational tricks used to get around the limitations of problems with the current framework, but this could easily have been building another workaround on a workaround without considering the mathematics involved.

When discussing the example of solving a variational problem, due to the marking of boundary points as constrained, the code in Firedrake is no longer taking in the initial guess and then modifying the guess so that it satisfies the Dirichlet boundary conditions imposed by the problem. Instead, it is now the default to leave out all the nodes of $u$ for which we know the value, and solve with $v$ for $u_h$ with both functions expanded through the nodal basis for $V_0$, even though the same output is produced. This is exactly what goes on when solving the matrix-vector system in Equation \ref{matrix-vector-system} introduced in the section on the mathematical theory of a finite element method. In contrast, when looking at how Firedrake previously treated boundary conditions, we noted that we solve with $v$ for $u_h$ with both functions expanded through the nodal basis for $V$, including the basis functions that correspond to the Dirichlet boundary conditions. Therefore, the \texttt{RestrictedFunctionSpace} implementation is much closer to the mathematics as described than the previous implementation and as such, the new class means that the algorithms utilised in Firedrake are closer to the mathematical ideas of Section \ref{section-fem}.

\newpage
\section{Appendix A: Consistency of a Mesh Numbering in Parallel}
\label{section-appendix-a}
In Section 5.6, we looked at a mesh numbering in parallel using a \texttt{RestrictedFunctionSpace} consisting of a degree 2 Lagrange element and 2 Dirichlet boundary conditions on $x = 0$ and $x = 1$. The numbering of the mesh points, and whether interpolating the functions $x$ and $y$ using the \texttt{RestrictedFunctionSpace} were successful, was given in Figure \ref{numbering-issues-parallel}. Here, we construct a table for Process 0 to show each mesh point, its offset and the corresponding \texttt{local\_to\_global\_map} and \texttt{plex\_renumbering} output. It should be noted that the \texttt{local\_to\_global\_map} is calculated using the offsets, and the \texttt{plex\_renumbering} output is calculated with the mesh points.
\begin{table}[h!]
    \centering
\begin{tabular}{|c|c|c|c|c|c|}
\hline
    Mesh Point & PyOP2 Label & Offset & LGMap Index & Plex Renumbering Index & Constrained? \\ \hline 
     0 & Owned & 0 & N/A & 0 & No (N/A) \\ 
     1 & Ghost & 3 & N/A & 4 & No (N/A) \\
     2 & Owned & 2 & -1 & 3 & Yes\\
     3 & Ghost & 6 & -1 & 8 & Yes \\
     4 & Ghost & 8 & -1 & 10 & Yes\\
     5 & Ghost & 7 & -1 & 9 & Yes\\
     6 & Owned & 1 & -1 & 2 & Yes\\
     7 & Owned & 0 & 0 & 1 & No \\
     8 & Ghost & 5 & 2 & 7 & No \\
     9 & Ghost & 3 & 1 & 5 & No \\ 
     10 & Ghost & 4 & -1 & 6 & Yes \\
     \hline
\end{tabular}
    \caption{Table of the mesh points as shown in Figure \ref{numbering-issues-parallel} for Process 0, their PyOP2 label, their offset and their respective index in the \texttt{LGMap} and plex\_renumbering}
    \label{table-mesh-properties}
\end{table} \par 
From Table \ref{table-mesh-properties}, we can look at the rows to see if the numbering of the mesh points and their offsets is consistent, when comparing to a Hasse diagram or similar of the mesh. The first thing that can be noted is that the \texttt{LGMap} only has non-negative indices when on non-constrained mesh points, which is what is expected. It is also possible to see that the plex renumbering returns higher indices for ghost points, and that constrained owned points 2 and 6 are behind the unconstrained owned point 7 (and also technically 0) in the plex renumbering order. Process 1 follows a similar pattern, meaning that no obvious issues have appeared.

The outcome of this analysis is that there are no immediate inconsistencies in the numbering, and the reordering of the mesh points through the plex renumbering has worked as expected. 

\newpage
\pagenumbering{gobble}

\bibliographystyle{unsrt}
\bibliography{references}

\end{document}